\documentclass[sigconf]{acmart}

\usepackage[most]{tcolorbox}
\usepackage{amsthm,amsmath}
\usepackage{xspace}
\usepackage{float}
\usepackage{circledsteps}
\usepackage{multirow}
\usepackage{wasysym}
\usepackage[table]{xcolor}
\usepackage{fontawesome5}

\usepackage[
n,
operators,
sets,
adversary,
landau,
probability,
notions,
logic,
ff,
mm,
primitives,
events,
complexity,
asymptotics,
keys]{cryptocode}

\usepackage[capitalise,noabbrev,nameinlink]{cleveref}
\usepackage{placeins}

\newcommand{\definedterm}{} 
\newtheorem*{worddefinner}{\definedterm}

\newcommand{\pufQ}{\mathcal{Q}}
\newcommand{\mathbbm}[1]{\text{\usefont{U}{bbm}{m}{n}#1}} 

\newcommand{\securityparam}{\ensuremath{{\lambda}}}

\ifdefined\negl
\renewcommand{\negl}{\ensuremath{\mathsf{negl(\securityparam)}}}
\else
\newcommand{\negl}{\ensuremath{\mathsf{negl(\securityparam)}}}
\fi

\newcommand{\targetingfunction}{\ensuremath{f_{\mathsf{t}}}\xspace}
\newcommand{\lensfunction}{\ensuremath{\rho}\xspace}
\newcommand{\reportcreationfunction}{\ensuremath{f_{\mathsf{r}}}\xspace}
\newcommand{\attributionfunction}{\ensuremath{f_{\mathsf{a}}}\xspace}
\newcommand{\engagementfunction}{\ensuremath{f_{\mathsf{e}}}\xspace}
\newcommand{\browsingfunction}{\ensuremath{f_{\mathsf{b}}}\xspace}

\newcommand{\randomchoice}{\ensuremath{\xleftarrow{\$}}}
\renewcommand{\sample}{\randomchoice}

\theoremstyle{definition}
\newtheorem{definition}{Definition}[section]

\newtheorem{theorem}{Theorem}

\mathchardef\mhyphen="2D

\newcommand{\advantage}[2]{\ensuremath{\mathsf{Adv}^{#1}_{#2}}\xspace}

\newcommand{\Done}{\mathcal{D}_1}
\newcommand{\Dzero}{\mathcal{D}_0}
\newcommand{\close}{\Delta}

\usepackage[shortlabels]{enumitem}

\newcommand{\idealtargeting}[0]{\ensuremath{\mathcal{F}_{Targeting}}\xspace}
\newcommand{\idealtargetingfull}[0]{\ensuremath{\idealtargeting^{\targetingfunction,\lensfunction}}\xspace}

\newcommand{\idealuser}[0]{\ensuremath{\mathcal{F}_{UserData}}\xspace}
\newcommand{\idealuserfull}[0]{\idealuser}
\newcommand{\idealsociety}[0]{\ensuremath{\mathcal{F}_{Society}}\xspace}
\newcommand{\idealsocietyfull}[0]{\ensuremath{\idealsociety^{n,\mathcal{D}}}\xspace}
\newcommand{\idealmetrics}[0]{\ensuremath{\mathcal{F}_{Metrics}}\xspace}
\newcommand{\idealmetricsfull}[0]{\ensuremath{\idealmetrics^{\attributionfunction,\reportcreationfunction}}\xspace}

\newcommand{\Adv}[0]{\ensuremath{Env}\xspace}
\newcommand{\Env}[0]{\ensuremath{Env}\xspace}

\newcommand{\idealengagement}[0]{\ensuremath{\mathcal{F}_{Engagement}}\xspace}
\newcommand{\idealengagementfull}[0]{\ensuremath{\idealengagement^{\browsingfunction,\engagementfunction}}\xspace}

\graphicspath{{graphics/}}

\AtBeginDocument{%
  }

\setcopyright{none}
\renewcommand\footnotetextcopyrightpermission[1]{}

\begin{document}

\title[Making Sense of Private Advertising]{Making Sense of Private Advertising: A Principled \\ Approach to a Complex Ecosystem}

\author{Kyle Hogan}
\orcid{0009-0001-3648-6319}
\affiliation{
\institution{Massachusetts Institute of Technology}
  \city{} 
  \country{} 
}
\email{klhogan@csail.mit.edu}

\author{Alishah Chator}
\affiliation{
 \institution{Baruch College}
 \city{}
 \country{}}
\email{alishah.chator@baruch.cuny.edu}
\authornote{Work done while at Boston University}

\author{Gabriel Kaptchuk\footnotemark[\value{footnote}]}
\affiliation{
 \institution{University of Maryland}
 \city{}
 \country{}}
\email{kaptchuk@umd.edu}

\author{Mayank Varia}
\affiliation{
 \institution{Boston University}
 \city{}
 \country{}}
\email{varia@bu.edu}

\author{Srinivas Devadas}
\affiliation{
 \institution{Massachusetts Institute of Technology}
 \city{}
 \country{}}
\email{devadas@mit.edu}

\renewcommand{\shortauthors}{Hogan et al.}

\begin{abstract}
    In this work, we model the end-to-end pipeline of the advertising ecosystem, allowing us to identify two main issues with the current trajectory of private advertising proposals.
First, prior work has largely considered ad targeting and engagement metrics individually rather than in composition. 
This has resulted in privacy notions that, while reasonable for each protocol in isolation, fail to compose to a natural notion of privacy for the ecosystem as a whole, permitting advertisers to extract new information about the audience of their advertisements. 
The second issue serves to explain the first: we prove that \textit{perfect} privacy is impossible for any, even minimally, useful advertising ecosystem, due to the advertisers' expectation of conducting market research on the results.

Having demonstrated that leakage is inherent in advertising, we re-examine what privacy could realistically mean in advertising, building on the well-established notion of \textit{sensitive} data in a specific context.  We identify that fundamentally new approaches are needed when designing privacy-preserving advertising subsystems in order to ensure that the privacy properties of the end-to-end advertising system are well aligned with people's privacy desires.

\end{abstract}


\maketitle
\pagestyle{plain}

\section{Introduction}
Behavioral advertising, in which people are preferentially shown advertisements that align with their interests and demographics, has become the financial backbone of the internet and the default business model for large swaths of the technology sector.
This advertising ecosystem---and the user-tracking infrastructure that powers it---are widely known to be privacy invasive \cite{pewads}, not only due to the collection of sensitive, personal data, but also because of the ways in which that data is \textit{used}~\cite{NDSS:DMYZG16,USENIX:CabCueCue18,SP:VALMGL18, PoPETS:VLSM19}.

A great deal of research has focused on the harms caused to people through the use of their data for the targeting of advertisements. 
Targeting algorithms are designed to maximize profit for the constituents of the advertising industry, not to serve the best interests of the viewer, and prior studies have shown that behavioral targeting is used to perpetuate harmful biases \cite{imana2021auditing,datta2018discrimination,speicher2018potential,203840,291188}, facilitate the spread of disinformation \cite{wiredmisinformation,wood2017fool,crain2019political,hiltunen2021online}, and exploit sensitive information such as mental health data to target vulnerable populations \cite{USENIX:CabCueCue18,carrascosa2015always,wei2020twitter,markup2023audiences}.

While targeting has been the most widely-researched component of behavioral advertising, matching ads to users is not the only functionality of the ecosystem.
Advertisers expect to conduct market research using the relative success of their different advertisements.
To do so, they demand \textit{metrics} on how users responded to ads---did they buy something after viewing the ad? Subscribe to the advertiser’s mailing list? If so, which ad drove this engagement?
Ad networks keep records of this information and submit it back to advertisers, allowing them to improve their understanding about the preferences of their consumers and refine future ad campaigns.

Targeted advertising is not unique to digital advertising: print advertisements have long been targeting their ads towards specific groups of people using both the context in which the ad will be displayed (e.g.,
the
magazine or billboard on which to advertise) as well as cues within the ad material itself (e.g., visuals, audio, etc.).  A famous example is the 90's Subaru campaign that was targeted towards the LGBT community by including, e.g., queer-coded license plates on the depicted cars \cite{subaru2002,mayyasi2016subarus}.
The shift to digital advertising is, therefore, not a change in \emph{type}, but a change in \emph{magnitude} and \emph{precision}. Digital advertising, with its specific target audiences and accurate attribution of user behavior to associated ad views,
allows advertisers to conduct market research that
is far more invasive than was possible with print media, calling into question the ethics of campaigns that focus on sensitive audiences. 
This is the case not only due to the well-understood harms of collecting sensitive data for targeting ads \cite{sherman2021data}, but also because it is currently unclear what or how much information advertisers are able to learn about these sensitive audiences from the metrics released on their ad campaigns.

In response, researchers and technology companies have proposed a shift towards ``privacy-preserving advertising systems,'' a collection of proposals \cite{zhong2022ibex,zhong2023addax,themis,ppad,adscale,obliviad,privad,adnostic,juels2001targeted,adsplit,ahead,badass,adveil,tholoniat2024cookie,ara,pcm,ipa} that aim to maintain the existing advertising business model while making the process of (1) targeting advertisements and (2) reporting metrics on advertisement efficacy in a \textit{privacy-preserving} manner.  Proponents of privacy-preserving advertising systems have made significant progress in refining the design of these systems, some of which have even been deployed in popular browsers \cite{ara,mozilla-deployment,pcm}.

\vspace{0.5mm}
\noindent
\textbf{Unpacking ``privacy'' within advertising.}
In this work, we take a step back to re-examine ``privacy-preserving advertising systems'' from first principles.  This choice is motivated by the desire to better understand what is possible to achieve when adding privacy to advertising when, largely, the concept feels like an oxymoron.

Ultimately, we find that this initial reaction is not far off: any 
\textit{useful} behavioral advertising ecosystem must necessarily permit advertisers to extract information about end users, regardless of what privacy protections are in place. 
Leveraging the language of ideal functionalities, we give an implementation-independent modeling of privacy-preserving advertising that focuses on the \textit{minimal} functionality required by the ecosystem.

Notably, our model departs from prior work in that it focuses on the entire \textit{end-to-end} advertising pipeline, with a particular emphasis on the ways that privacy-preserving targeting and privacy-preserving metrics interact with one another and form a feedback loop.
By seeing privacy-preserving advertising in this way, we are able to identify real-world advertising use cases in which common notions of privacy for targeting and metrics fail to compose satisfyingly, which undermines natural privacy guarantees for the end-to-end system despite targeting and metrics protocols \textit{independently} achieving reasonable standards for privacy. 
We emphasize that this composition failure is not merely the result of specific, ill-designed protocols from early research; instead, it is fundamental to the nature of targeted advertising itself.

Looking further: we also examine how to consider privacy formally in the context of advertising.
While \textit{perfect} privacy may not be possible for advertising, we observe that it is also not necessarily required or even desirable. 
For instance, while contextual advertising (where ads are targeted only to the context in which they will be displayed, i.e., the website or article) can suffer from the same problems as behavioral advertising in the worst case,
it is still strongly preferred as an alternative to behavioral advertising. While the privacy provided by contextual advertising may be imperfect, it is likely ``good enough'' \cite{cdtcontextual,nytcontextual}.
Hence, using the language of information leakage alone makes it difficult to distinguish between advertising systems that are widely considered ``invasive'' and those that are not. 
We additionally observe that a narrow focus on building mechanisms to regulate---without eliminating---information leakage, like differential privacy \cite{TCC:DMNS06}, risk treating all types of information leakage identically and missing the ways in which \emph{people} feel differently about some sensitive information categories.
As a result, we adopt the framework of \emph{attribute privacy} \cite{zhang2022attribute} to evaluate privacy in the context of advertising.

Relatedly, not all advertising campaigns have the same potential for privacy harm, even if they do leak the same amount of information.
For this reason, we also employ the framework of contextual integrity \cite{nissenbaum2004privacy} to reason about the \emph{sensitivity} of the data involved.
This sentiment is captured, if not well-enforced, by current tech policy and legal regulations for ad targeting, but absent from consideration in private metrics.
This motivates a more holistic approach
toward the design of
privacy-preserving advertising systems that reason carefully about the amount of information revealed by metrics \emph{and} the sensitivity of that leakage.

\subsection{Our Contributions}
In this work, we provide a careful accounting of the structure underpinning the advertising ecosystem.  In doing so, we make the following contributions:
\begin{itemize}[--,leftmargin=*,itemsep=0pt]
    \item \textbf{A clean, formal, and flexible abstraction of the end-to-end advertising process.}  
    Our model, detailed in \cref{sec:model}, makes extensive use of parameterizing functions to ensure that our abstraction is flexible enough to describe the behavior of the real-world advertising pipeline as well as ongoing proposals to make it more private.  We choose Canetti's Universally Composable (UC) security framework\footnote{As we discuss in \cref{sec:model}, the way in which we use this model is non-standard, as we use it to prove \emph{a lack of security}.} \cite{FOCS:Canetti01} as the runtime for our modeling in order to give structure to our analysis. 

    \item \textbf{A formal illustration of the inherent tension between privacy and utility.}  In \cref{sec:defining-privacy}, we highlight a gap between the individual focus of current private advertising protocols and the structure of advertising itself which considers \textit{audiences}, or groups of people.
    Leveraging our model, we concretize this gap in \cref{sec:privacyvsusefulness} by providing lightweight minimum utility notions required of each component of an advertising ecosystem, which we use to prove an inherent incompatibility with the group privacy notion, \textit{attribute privacy} \cite{zhang2022attribute}.
    Specifically, we prove that any \textit{useful} advertising ecosystem must necessarily leak some information about its users---even when it employs strong, individual privacy protections, such as differential privacy.
    We characterize this leakage both theoretically and empirically in terms of the difference in \emph{sample complexity}~\cite{canonne2020survey}---the campaign size required for a private advertising ecosystem to leak the same information as its non-private counterpart.

    \item \textbf{A refocus on normative privacy notations.} Having shown that some
    leakage is inherent in advertising, in \cref{sec:redefining-privacy} we advocate for rooting future discussions of privacy in data \textit{sensitivity} as understood by end users, ad tech platforms, and regulatory bodies.
    Data sensitivity is well-studied when it comes to private ad targeting, but is largely ignored by private metrics protocols, which focus exclusively on the \textit{quantity} of leakage.
    We propose that by making metrics \textit{targeting-aware}, protocols could incorporate the idea of sensitivity and serve as a second layer of enforcement--and accountability--for private advertising, refocusing on \textit{what} information is revealed about users.

\end{itemize}

\section{Background on (Private) Advertising}
\label{sec:background}

\subsection{The Advertising Ecosystem}
\label{sec:background:advertising}

\begin{figure*}[ht]

\includegraphics[width=\textwidth]{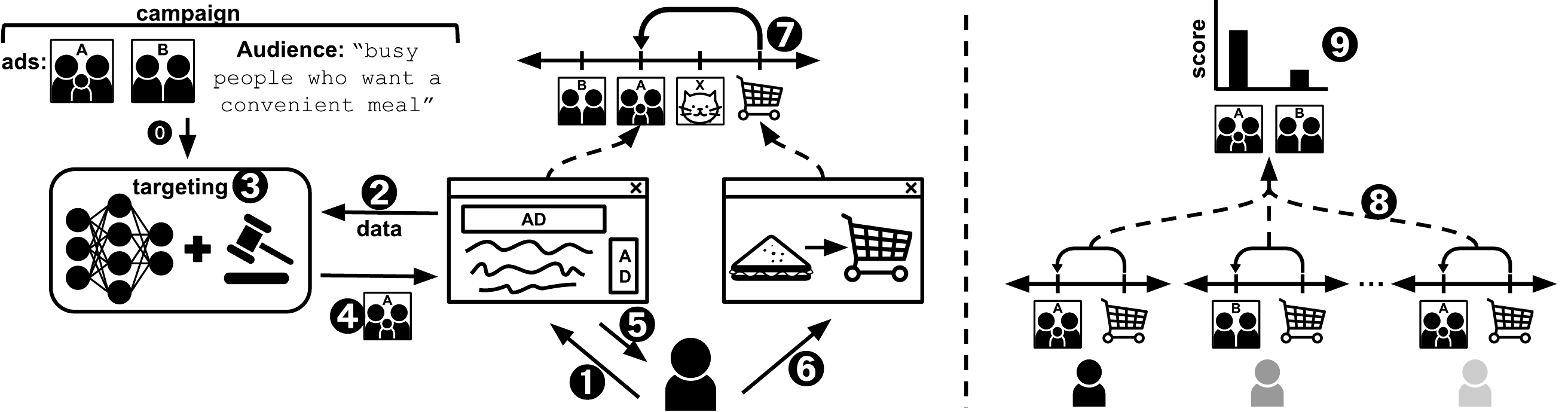}
\caption{An illustration of the advertising ecosystem, depicting the process of generating metrics on a behavioral advertising campaign.  See \cref{sec:background:advertising} for details.}
\label{fig:adsecosystem}
\end{figure*}

The language used within the advertising literature can be difficult to parse for the unfamiliar reader.  As such, we provide a brief overview of digital advertising, focusing on the creation of metrics data (we
direct the reader to other surveys for more detail \cite{pooranian2021online}).

We illustrate the life cycle of an advertising campaign in \cref{fig:adsecosystem}. 
A \textit{campaign}, or collection of ads directed at a target audience, begins at step \Circled{0} when the advertiser registers it with the ad network. 
In this case, the advertiser is using their campaign to conduct an A/B test, a common practice that we discuss in depth in \cref{sec:defining-privacy}, by registering two ads: one ad (A) depicting a family with children and another ad (B) without children. In both cases the ads are directed at an audience of ``busy people looking for a convenient meal''. 

Later, \Circled{1} when a user browses to a publisher website displaying ads, that website will send a request \Circled{2} to the ad network containing data on the site itself as well as an identifier for the user. The ad network then runs a targeting model and a real-time bidding auction \Circled{3}, incorporating data from the advertiser, in order to select which ad to deliver to the user \Circled{4}. The user views the ad \Circled{5} and may, at some point in the future \Circled{6} subscribe to a meal kit service (likely on a different site) as a result of seeing this ad. 

Engagement with an advertiser, such as making a purchase or even adding items to a cart, is known as a \textit{conversion}.
The ad network attributes \Circled{7} the user's conversion to the \textit{impression}, or ad view, it believes was responsible. In this case, that was the most recent impression, ignoring ads from unrelated campaigns. 
This process is known as \textit{attribution}, and attributing a conversion to the most recent impression is a strategy called ``last touch.''

Eventually, the ad network collects attribution data \Circled{8} from \textbf{all} users who viewed ads in this campaign and uses it to compute metrics \Circled{9}
that
the advertiser
can use to determine
which ad from the campaign was more successful in driving purchases.

In more detail, these metrics largely correspond to a count of how many conversions were attributed to each ad, and they are what allow the advertiser to run its A/B test and refine their strategy for future campaigns
to focus
more
effort on ad content that drove higher engagement \cite{johnson2023inferno,braun2024leveraging}.
For this example, ad A outperformed ad B, so
the advertiser will likely focus future advertising spend towards parents.
We go into more detail on the leakage from advertising metrics and its potential for harm in \cref{sec:defining-privacy}.

\subsection{Related Work}

While there is a rich history of academic research on privatizing advertising \cite{zhong2022ibex,zhong2023addax,themis,ppad,adscale,obliviad,privad,adnostic,juels2001targeted,adsplit,ahead,badass,adveil,tholoniat2024cookie,liu2016privacy}, the majority of this work
has not considered the potential impact of releasing metrics data (beyond the possibility of linking an individual conversion report to the specific user who generated it).
Exceptions to this include
AdVeil \cite{adveil},
Themis \cite{themis},
CookieMonster \cite{tholoniat2024cookie},
and various industry proposals for privatizing metrics (Apple's PCM/PAM \cite{pcm,pam}, Google's ARA \cite{ara}, Meta/Mozilla's IPA \cite{ipa}, and a W3C standardization effort PPA \cite{ppa}) that we discuss next.

Adveil \cite{adveil} presents a protocol for the full advertising pipeline, and
it considers the fact
that metrics reports can be revealing of personal data even if they are not directly linkable to an individual user.
Themis \cite{themis} is an early industry proposal that uses a consortium blockchain to provide transparency and accountability for metrics data. It, again, provides unlinkability between users and their reports, but it does not have further privacy protections for metrics data.
CookieMonster \cite{tholoniat2024cookie} is a recent work out of the W3C Private Advertising Technology Community Group (PATCG), which is working to standardize a private metrics protocol.
CookieMonster
provides a full model and security analysis for differentially-private ad metrics with emphasis on handling complexities in privacy loss budgets.
It
grew out of earlier work on Interoperable Private Attribution (IPA) \cite{ipa} and is part of the work on Privacy Preserving Attribution (PPA) \cite{ppa}.
Apple's Private Attribution Measurement (PAM) \cite{pcm} and Google's Attribution Reporting API (ARA) \cite{ara} are both alternative, differential-privacy-based proposals---though PAM was recently superseded by PPA. 

To the best of our knowledge, ours is the first work to formally model and prove the \textit{presence} of leakage for \textit{all} advertising systems, rather than the absence of leakage for a specific system.

\section{Defining Privacy for Advertising}
\label{sec:defining-privacy}

Privacy is a multifaceted \cite{solove2005taxonomy} and contextually embedded \cite{mulligan2016privacy} concept that does not permit a unified definition, so we first concretize what we mean by privacy within advertising.  
We begin with a \emph{leakage}-based notion; ideally, advertising systems that aim to preserve privacy should prevent \emph{any} information from leaking about users. 
Emerging proposals for private advertising are rapidly moving towards this ``no-leakage'' world by pushing more of the targeting logic to clients' own devices (e.g., FLEDGE \cite{fledge}).
Such a shift is a step in the right direction---away from the mass surveillance~\cite{christl2017corporate,buchi2022chilling,segijn2023ethical,yun2020challenges,van2020personalization} that currently supplies personal data for ad targeting.
However, delivering relevant ads is only one step in the advertising pipeline.

Advertisers also want metrics on how these ads perform. Ads can be expensive, and performance metrics allow
advertisers
to direct their spending to campaigns that drive a better return on ad spend (ROAS) \cite{pooranian2021online}. 
Yet, even very basic metrics, such as which ads were delivered, violate our zero-leakage goals as, due to the nature of targeting, the ads themselves are revealing of their audience \cite{meng2016price,castelluccia2012betrayed}.
Recognizing that metrics leakage could cause significant harm, there has been a widespread effort~\cite{ppa,tholoniat2024cookie} to make the metrics computed on advertisement performance differentially private, limiting the amount of information contained about individuals. However, as we demonstrate in this work, advertising requires more than individual privacy in order to adequately protect the information that users consider to be important.

We use this section to introduce metrics as a critical component of the advertising ecosystem, outline why it renders perfect privacy impossible for advertising, and argue that seemingly-natural fixes---such as differentially-private metrics---fail to adequately mitigate the privacy harms that can arise from advertising ecosystems.

\subsection{Market Research as Information Leakage}

The insights that advertisers derive from metrics go far beyond simple counts of how often advertisements are shown, and
they
are used to conduct market research on how users engage with the ads they are shown. 
By collecting metrics on the relative successes of their current advertising campaigns, advertisers can refine the content and target audience of future campaigns to focus their ad spending on serving appealing content to the people who are most likely to engage with it \cite{GOLDFARB2011289}.  

The most clear example of this type of market research is the A/B test, a practice where advertisers can create two versions of an ad and test which is preferred by their target audience or, conversely, test which of two possible target audiences gets better results for a given ad campaign.
A/B tests are so commonplace that major advertising platforms have built-in tools for advertisers to set up their experiments.\footnote{See, for example, \href{https://www.facebook.com/business/ads/ab-testing/}{Facebook A/B Testing} and \href{https://support.google.com/admanager/answer/7661678?hl=en}{Google A/B Testing}.}

To illustrate how such tests are conducted, we revisit \cref{fig:adsecosystem} and consider an instant meal-kit company that wants to decide whether to focus its ad spend toward parents with young children.
Such a company could set up its campaign in two main ways, testing on the target audience or on the ad content:
\begin{enumerate}[left=0pt,label=(\arabic*),itemsep=0pt]
    \item Create two different ads, both depicting someone in a rush using the meal-kit to prepare a quick meal, but $\texttt{ad}_A$ features a toddler and $\texttt{ad}_B$ does not. 
    \item Show the generic meal-kit ad (i.e., one without any particular features that suggest its relevance to parents) to two different audiences, audience A being, e.g., ``busy parents looking for a convenient meal option''\footnote{While this may seem quite abstract, in practice, targeting on these types of audiences is enabled with a combination of machine learning models, externally gathered data, and identification of lookalike audiences. See \href{https://help.criteo.com/kb/guide/en/about-audiences-0EJNOqUqYu/Steps/842036}{Criteo's Audience Overview} for more about audience generation practices.} while audience B removes the parent feature, e.g., ``people looking for a convenient meal option.''  
\end{enumerate}

No matter which A/B testing approach is leveraged, the result is fundamentally the same: advertisers learn whether members of their audience are more likely to be parents with young children based on the relative performance of A and B.
In case (1), this follows from one of the core axioms of advertising: people are more likely to engage with ads that are more relevant to them~\cite{kaspar2019personally,matz2017psychological}.
By contrast, in case (2)
the targeting algorithm
will preferentially show the ad it believes to be more relevant to the audience, i.e., if many of the audience members are parents, then ad A will be shown more frequently.

Market research is an iterative process.  Consider our earlier campaign example of marketing ready-made meal kits: 
the first iteration
could test
whether the audience of ``busy people who don't cook'' also tends to have the ``new parent'' feature.
Supposing that this turns out to be the case, the advertiser can then test whether ``busy new parents who don't cook'' tend to prefer ``health-focused'' meals and so forth.  Thus, this practice does not only reveal a ``little bit more'' information about audiences, but can be used (over time) to extract tremendous amounts of information about audiences.

Much market research is, like this example, relatively innocuous.
However, this same infrastructure can be (and is) used to learn about \textit{arbitrary} topics.
We see examples of this with researchers leveraging these platforms to carry out their own research studies \cite{seeman2024between}---studies that are suspiciously close to ``human subject research'' that is generally expected to be under the close supervision of institutional review boards. For instance, consider the study by Chan et al. that uses advertising to assess whether conservatives are likely to have stronger brand attachment \cite{chan2019political}.
Another study used advertising to assess public perceptions of refugees, though it did acknowledge the ethical considerations of the research \cite{adida2022refugees}. 

The upshot is that market research\textemdash an inherently desired component within any advertising system\textemdash enables a level of data-mining that goes far beyond improving the quality of advertising. However, the existing discussion of privacy in advertising is centered on \textit{individual} privacy, whereas
the leakage we describe here is a \textit{group} privacy harm.

\subsection{Distributional Privacy for Advertising}\label{sec:defining-privacy:nontrivial}

At first glance, this may seem like a natural place to utilize differential privacy (DP) \cite{TCC:DMNS06}, and indeed most proposals for private advertising metrics systems use DP. However, simply privatizing the aggregated metrics in this way is insufficient. Market research involves inference over a target audience or \textit{group} of people, not an individual user, and DP is intentionally designed to enable this type of inference~\cite{dwork2013s}.

Unlike a typical research study, the selection process for advertising audiences is designed to ensure that their members are \textit{not} representative of the general population \cite{chouaki2022exploring,korolova2010privacy}. 
Instead, audiences will often overwhelmingly represent small, arcane minority groups whose members may not even realize that such a grouping exists \cite{miller2025invisible}: examples of audience profiles include ``receptive to emotional messaging,'' ``rollercoaster romantics,'' ``heavy buyers of pregnancy tests,'' and ``strugglers and strivers -- credit reliant'' \cite{markup2023audiences}.

Many of these audiences represent vulnerable populations and allowing advertisers to extract arbitrary information about them can be harmful even when it doesn't permit linking this information to individuals. Manipulative advertising practices use these inferences to tailor their messaging to the viewer, increasing its effectiveness \cite{Borenstein02042024,zard2023consumermanipulationonlinebehavioral,susser2019online}. A common example of this practice is in political advertising where ads are typically \textit{microtargeted} to specific populations with the intent to influence their vote \cite{zuiderveen2018online,ribeiro2019microtargeting,Kruikemeier2015political}.

An additional challenge is that DP doesn't provide protection against an advertiser applying the group-level inferences made over the audience to its individual members \cite{seeman2024between}.
A common example for DP is that learning the group trend that ``smoking causes cancer'' would also imply that any specific smoker is at risk of cancer.
But advertising takes this a step further due to
``custom audiences'' that can be composed of specific, identifiable individuals such as those on the advertiser's mailing list.
For these audiences, a better analogy might be revealing that members of a specific sci-fi book club have an unusually high rate of cancer.
Unlike in the case of smoking, this is not a global inference implying that sci-fi books cause cancer, but is instead reflective of the health status of these specific people.
Revealing this type of inference---even with DP guarantees---is likely counter to peoples' expectation of privacy, especially given the information asymmetry in advertising where inferences are revealed to the advertisers, but not their audience.
For this reason, we instead employ \textit{attribute privacy} \cite{zhang2022attribute} to capture the potential privacy harms from advertising market research.

\subsection{Attribute Privacy in the Advertising Context}
\label{sec:attributeprivacy}
Attribute privacy, proposed by \citet{zhang2022attribute}, describes the ability of an adversary to learn information about specific, sensitive attributes of a population given summary statistics about that population.
It defines sensitivity around the maximum contribution that the \textit{distribution} of some sensitive feature in a population may have on the output of statistics computed over that population.

In this section, we provide the formal definition of attribute privacy, which is built on the pufferfish privacy framework \cite{kifer2014pufferfish}.\footnote{For some background on Pufferfish privacy, see \cref{app:pufferfish}.}
Later, in \cref{sec:redefining-privacy}, we provide some guidance on how attribute privacy could be integrated into the advertising ecosystem as a potential enforcement mechanism for user-focused privacy policies.

\begin{definition}[Dataset Attribute Privacy, Definition 3 from \citet{zhang2022attribute}]
\label{defn:datasetattributeprivacy}
Let $(X_1^j, X_2^j,\ldots,X_m^j)$ be a record with $m$ attributes that is sampled from an unknown distribution $\mathcal{D}$, and let $X=[X_1,\ldots,X_m]$ be a dataset of $n$ records sampled i.i.d.~from $\mathcal{D}$ where $X_i$ denotes the (column) vector containing values of the $i$th attribute of every record. Let $C\subseteq [m]$ be the set of indices of sensitive attributes, and for each $i \in C$, let $g_i(X_i)$ be a function with codomain $\mathcal{U}^i$.

A mechanism $\mathcal{M}$ satisfies \emph{$(\epsilon,\delta)$-dataset attribute privacy} if it is $(\epsilon,\delta)$-Pufferfish private for the following framework $(S,\pufQ,\Theta)$:
\begin{description}
\item Set of secrets: $S=\{s^i_a := \mathbbm{1}[g_i(X_i) \in \mathcal{U}^i_a] : \mathcal{U}^i_a \subseteq \mathcal{U}^i, i\in C\}$.
\item Set of secret pairs: $\pufQ=\{(s_a^i,s_b^i) \in S \times S, i\in C\}$.
\item Distribution: $\Theta$ is a set of possible distributions $\theta$ over the dataset $X$. For each possible distribution $\mathcal{D}$ over records, there exists a $\theta_{\mathcal{D}} \in \Theta$ that corresponds to the distribution over $n$ i.i.d.~samples from $\mathcal{D}$.
\end{description}
\label{def:dbprivacy}
\end{definition}

To contextualize this definition in the advertising setting, consider the dataset $X$ to be an advertising audience with $n$ members, each represented by a feature vector $(X_1^j, X_2^j,\ldots,X_m^j)$ of length $m$ indicating the attributes of the $j^{th}$ user. Some attributes will be considered \textit{sensitive} and represented in $C\subseteq [m]$.\footnote{In \cref{sec:redefining-privacy}, we employ the contextual integrity framework \cite{nissenbaum2004privacy} to provide guidance on how to decide which attributes might be sensitive in the context of advertising.} Then:
\begin{itemize}[left=0pt]
\item The secret pairs $(s_a^i,s_b^i)$ for a sensitive attribute $i$ are possible realizations of some function $g_i(X_i)$ over that sensitive attribute. For advertising metrics, we can think of $g_i()$ as computing the fraction of the audience who possess the sensitive attribute.

\item $\Theta$ is the set of possible distributions that could have generated the audience shown in $X$. Each $\theta$ is intended to capture possible correlations across attributes.
\end{itemize}

Formally, the sensitivity of an output statistic $F(X)$ over the dataset $X$ is computed as follows:
\begin{equation}\label{def:gaussian_sens}
\Delta_i F = \max_{\theta\in \Theta}\max_{(s_a^i,s_b^i)\in \pufQ}\abs{\mathbb{E}[{F(X)|s_a^i,\theta}]-\mathbb{E}[{F(X)|s_b^i,\theta}]}.
\end{equation}
For advertising, we consider $F(X)$ to be the metrics for an advertising campaign (e.g., a count of ad clicks, conversions, or purchases).
Sensitivity captures the maximum impact on $F(X)$, which occurs for the
pair of potential secrets $(s_a^i,s_b^i)$
in which all or none of the audience (respectively) have the sensitive attribute and for the $\theta$ with the tightest correlation between
this
attribute and the conversion rate.
In words: if possessing the sensitive attribute makes a user significantly more (or less) likely to engage with an ad, then varying the prevalence of this feature within the audience will have a strong impact on reported number of conversions.
For instance, in our example A/B test from the previous section with an audience of ``busy people who don't cook,'' a toddler-focused ad is much more strongly correlated with the sensitive attribute (parental status) than an ad focused on the types of food contained in the meal kit.

We demonstrate
later in \cref{sec:privacyvsusefulness}
that a lack of attribute privacy is inherent to advertising; i.e.,
any minimally useful ads ecosystem will reveal some new information about its audiences.
However, current instantiations and associated privacy definitions give very little control over \textit{what} information leaks.
In \cref{sec:redefining-privacy}, we discuss the concept of sensitivity in depth and argue that advertising requires more than individual privacy in order to meet users' expectations.

\section{Modeling the Advertising Ecosystem}
\label{sec:model}

\begin{figure}[t]
\includegraphics[width=0.35\textwidth]{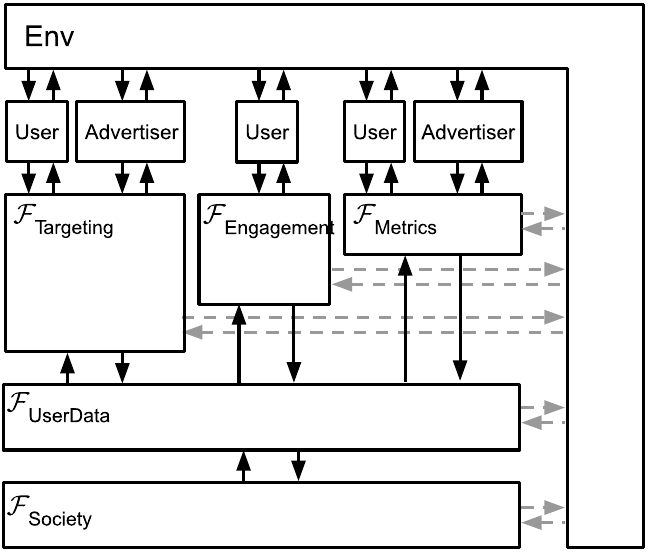}
\caption{UC functionalities for advertising ecosystem.}\label{fig:frameworkoverview}
\end{figure}

In this section, we present our minimalist modeling of the advertising ecosystem.  Our modeling captures the \emph{minimum} information leakage present in the advertising ecosystem, and  it represents an ecosystem that has been designed to eliminate all unintended information flows back to advertisers. We perform this modeling from the perspective of advertiser by having the advertiser set target audiences and receive summary reports on ad display and engagement. While other actors in the advertising ecosystem (e.g., publisher websites and ad networks) certainly have their own functionality goals, advertisers are the driving force behind the feedback loop on user data through the advertising ecosystem.  In \cref{sec:privacyvsusefulness} we will use this model to ``prove \emph{insecurity},'' i.e., show that useful advertising ecosystems will necessarily leak information about their users.  
Thus, if some future system provably instantiates our ideal functionalities, that should not be misconstrued as a demonstration that it is privacy-preserving in a normative sense. Instead, such a system would have \textit{at least} the leakage we demonstrate here, and may have substantially more.

\begin{figure}[t]
	\ifdefined\cryptolayout
	{\footnotesize
	\fi
\begin{center}
	\begin{tcolorbox}[enhanced,sharp corners,colback=white,boxrule=0.3mm,left=1mm]
		\vspace{0.5\baselineskip}
		Ideal Functionality $\idealsocietyfull$

		\vspace{0.5\baselineskip}
		\hrule 
		\vspace{0.5\baselineskip}

        \idealsociety is parameterized with a number of individuals $n$ and a distribution over the user feature space $\mathcal{D}.$

        \smallskip
        \textbf{Initialize:} Upon receiving an \texttt{init} message from $\idealuserfull$ for $User_i$:
        \begin{enumerate}[noitemsep]
            \item if $User_i$ does not already have recorded features, sample features for $User_i$ from $\mathcal{D}$ and record.
            \item Send an \texttt{init} message to \idealuser with $User_i$ and its recorded features.
        \end{enumerate}

	\end{tcolorbox}	
\end{center}
\ifdefined\cryptolayout
}
\vspace{-2em}
\fi
\caption{$\idealsocietyfull$, our way of modeling the features provided to people in society.}
\label{fig:society}
\ifdefined\cryptolayout
\vspace{-2em}
\fi
\end{figure}

\subsection{Parameterizing Functions}
\label{sec:parameterizingfunctions}
Our model makes heavy use of parameterizing functions when specifying ideal functionalities.   These parameterizing functions mean that our model is \emph{flexible} enough to capture a wide variety of potential advertising systems, including those that attempt to preserve privacy, those that are widely understood to be privacy invasive, or even systems that would make little sense to deploy in practice.  We note that it might be best to think of some of these functions as being \emph{stateful} (e.g., if a privacy budget must be managed over many queries); for simplicity, we do not explicitly manage state for these functions, but observe that it is trivial to modify our modeling to make them stateful.  We briefly introduce these parameterizing functions before presenting our formal model. 

\smallskip
\noindent
\textbf{The Targeting Filter $\rho$ and Targeting Function $\targetingfunction$.}  We model the decision to select which ad for a user $user_i$ when they visit a website as a two phase process:
(1)
from the set of $user_i$'s \texttt{features},
a subset \texttt{features'} is 
extracted using a (\emph{deterministic}) filter function $\rho$
and is then
(2) fed into an arbitrary (\emph{randomized}) selection  function $\targetingfunction$, the output of which is an advertisement.  The filter function $\rho$ can be thought of as a policy that limits the type of information that the targeting logic $\targetingfunction$ is allowed to access.
Real-world 
instantiations of
$\rho$
could model: (1)
intentional restrictions such as Google's Topics API \cite{topics}
and (2) unintentional inaccuracies in targeting profiles. Then,
$\targetingfunction$ could embed any ``secret sauce'' used by the advertising network to select the most effective advertisement to match to a user, including an opaque machine learning model or even one of the academic proposals for private ad targeting and auctions \cite{zhong2023addax,zhong2022ibex}.

\begin{figure*}[ht]
	{\small
\begin{center}
	\begin{tcolorbox}[enhanced,sharp corners,colback=white,boxrule=0.3mm,left=1mm]
		\vspace{0.5\baselineskip}
		Ideal Functionality $\idealuser$

		\vspace{0.5\baselineskip}
		\hrule 
		\vspace{0.5\baselineskip}

        \idealuser has a set of users $\{User_0, User_1, ..., User_n\}$ each with a set of \texttt{features} (specified by \idealsocietyfull) and a \texttt{browsing-history}. Additionally, for each $User_i,$ $\idealuser$ also maintains three indices into their \texttt{browsing-history:} $\texttt{targeting}_i$, $\texttt{engagement}_i$, and $\texttt{attribution}_i$.

    \smallskip
        \textbf{Browsing:} Upon receiving a \texttt{browsing} message from \idealengagementfull for $User_i$:
        \begin{enumerate}[noitemsep]
            \item If $User_i$ has no \texttt{features}, send an \texttt{init} message to \idealsocietyfull and record the response. Set $\texttt{targeting}_i$, $\texttt{engagement}_i$, and $\texttt{attribution}_i$ to \texttt{0}.
            \item Send $User_i$'s identity, \texttt{features}, and full \texttt{browsing-history} to \idealengagementfull.
            \item Upon receiving a response from \idealengagementfull with \texttt{site}, append  \texttt{site} to  the \texttt{browsing-history} for $User_i$. Then, send \texttt{ok} message to \idealengagementfull.
        \end{enumerate}
        
        \smallskip
        \textbf{Ad Targeting:} Upon receiving a \texttt{target} message from \idealtargetingfull for $User_i$:
        \begin{enumerate}[noitemsep,nolistsep]
            \item If $\texttt{targeting}_i=\texttt{null}$ (i.e., \texttt{browsing} has not been called for this user) respond \texttt{fail} to \idealtargetingfull.
            \item Send $User_i$'s identity, \texttt{features}, and $\texttt{browsing-history}[0:\texttt{targeting}_i]$ to \idealtargetingfull.
            \item Upon receiving a response from \idealtargetingfull for $User_i$ of the form ($\texttt{site}$, \texttt{ad}), if $\texttt{browsing-history}[\texttt{targeting}_i]$ contains an matching entry $\texttt{site}$, then overwrite the entry with (\texttt{site}, \texttt{ad}) and increment $\texttt{targeting}_i$. Then, send \texttt{ok} to \idealtargetingfull.
        \end{enumerate}

        \smallskip
        \textbf{Ad engagement:} Upon receiving an \texttt{engagement} message from \idealengagementfull for $User_i$:
        \begin{enumerate}[noitemsep,nolistsep]
            \item If $\texttt{engagement}_i = \texttt{targeting}_i$ (i.e., engagement decisions were made for all impressions), respond \texttt{fail} to \idealengagementfull.
            \item Respond to \idealengagementfull with $User_i$'s \texttt{features} and the tuple $ \texttt{browsing-history}[\texttt{engagement}_i] = (\texttt{site}, \texttt{ad}).$
            \item Upon receiving a response from \idealengagementfull with $\texttt{ad}$ and $\texttt{conversion}$ for $User_i$, overwrite $\texttt{browsing-history}[\texttt{engagement}_i]$ to be $(\texttt{site}, \texttt{ad}, \texttt{conversion})$ and increment $\texttt{engagement}_i.$ Then, send \texttt{ok} message to \idealengagementfull.
        \end{enumerate}

        \smallskip
        \textbf{Attribution:} Upon receiving a \texttt{attribute} message from \idealmetricsfull for $User_i$:
        \begin{enumerate}[noitemsep,nolistsep]
            \item Let $j = \texttt{attribution}_i.$
            Starting with $j$, find the first entry of \texttt{browsing-history} with a tuple $(\texttt{site}, \texttt{ad}, \texttt{conversion})$ such that \texttt{conversion} is non-\texttt{None}.  Set $\texttt{attribution}_i$ to be the index of this entry.  If no such index exists, respond \texttt{fail} to \idealmetricsfull instead.
            \item Send an \texttt{attribution} message to \idealmetrics for $User_i$ with $\texttt{browsing-history}[j:\texttt{attribution}_i].$
        \end{enumerate}

	\end{tcolorbox}	
\end{center}
\caption{\idealuser holds data about each user and is responsible to shuttling information between $\idealtargeting, \idealengagement,$ and $\idealmetrics$. Handles user features along with all data related to ad impressions and conversions.}
\label{fig:user}
}
\end{figure*}

\smallskip
\noindent
\textbf{The Browsing Function \browsingfunction and Engagement Function \engagementfunction.} We make use of two (randomized) parameterizing functions, \browsingfunction and \engagementfunction
in order
to capture \emph{human behavior} in our model.  Specifically, \browsingfunction decides which website a user $User_i$ will visit, and \engagementfunction decides how a user $User_i$ will interact when presented with a particular advertisement (e.g., will they generate a ``conversion,'' by purchasing the advertised product). These are best thought of as ``black boxes''
that
need not be opened in order to understand the end-to-end functioning of the system.

\smallskip
\noindent
\textbf{The Attribution Function \attributionfunction.} Within the advertising ecosystem, each \emph{conversion} event must be \emph{attributed} to an advertisement impression (see \cref{sec:background}).  The attribution function $\attributionfunction$ performs the logic of this attribution.  For example, a common attribution function is ``last-touch attribution,'' where the most recent impression prior to a conversion receives all the ``credit'' for the conversion.   In the name of generality, the parameterizing attribution function \attributionfunction takes in a set of impressions (along with the context in which the impression occurred) and allocates scores to each of these impressions according to arbitrary logic. In practice, \attributionfunction could be instantiated by the Privacy Preserving Attribution protocol \cite{ppa}.

\smallskip
\noindent
\textbf{The Reporting Function \reportcreationfunction.}  Advertisers learn about the performance of various advertisements within a campaign by generating a report.  The exact nature of how this report is compiled from the attribution scores is system specific, but we encourage readers who want a concrete example to think about the report as simply a histogram of advertisement performance (i.e., a measure of how efficiently advertisement impressions became conversions).  In many of the emerging private advertising system proposals \cite{ipa,ara,pam,ppa}, report generation is done with differential privacy. We capture this process generically with the \reportcreationfunction parameterizing function, which could be instantiated by any of these proposals or by some future protocol with an alternative privacy mechanism.

\begin{figure}[t]
\begin{center}
	\begin{tcolorbox}[enhanced,sharp corners,colback=white,boxrule=0.3mm,left=1mm]
		\vspace{0.5\baselineskip}
		Ideal Functionality $\idealtargetingfull$
		
		\vspace{0.5\baselineskip}
		\hrule 
		\vspace{0.5\baselineskip}

        \idealtargeting~ is parameterized by  a \emph{stateful}, randomized function $\targetingfunction$ and a filtering function $\lensfunction$. It also maintains a set \texttt{active-campaigns}.

        \smallskip
        \textbf{Register Campaign:} Upon receiving a $\texttt{campaign}:(\texttt{audience}, \{\texttt{ad}_1, \ldots, \texttt{ad}_k\})$ message from the \Adv:
        \begin{enumerate}[noitemsep]
            \item Add \texttt{campaign} to \texttt{active-campaigns}.
            \item Send an \texttt{ok} message to the \Adv.
        \end{enumerate}

        \smallskip
        \textbf{Target Ad:} Upon receiving an \texttt{ad} message from \Env for $User_i$:
        \begin{enumerate}[noitemsep,nolistsep]
            \item Send a \texttt{target} message to \idealuser for $User_i$.
            \item Upon receiving a response from \idealuser with \texttt{features} and \texttt{browsing-history} for $User_i$:
            \begin{enumerate}[noitemsep,nolistsep]
                \item run $\texttt{features'} \gets \lensfunction(User_i, \texttt{features})$ to obtain $\texttt{features'} \subseteq \texttt{features}$
                \item Extract \texttt{site} from the final element of \texttt{browsing-history}.
            \end{enumerate} 
            \item Compute $ \texttt{ad} \gets \targetingfunction$(\texttt{active-campaigns}, \texttt{features'}, \texttt{site}).
            \item Send \idealuser~ a message  with the identifier $User_i$ and a tuple of the form $(\texttt{site}, \texttt{ad})$.
            \item Upon receiving \texttt{ok} or \texttt{fail} from \idealuser, send \texttt{ok} to the \Adv.
        \end{enumerate}

	\end{tcolorbox}	
\end{center}
\caption{Targeting functionality $\idealtargeting$}
\label{fig:targeting}
\end{figure}

\subsection{Ideal Advertising Functionalities}
\label{sec:modeloverview}

Before providing an overview of our model, we first introduce definitions for an ad and an audience
(we represent the latter using
\textit{feature vectors}
in this work).

\begin{definition}[Ad] 
\label{defn:ad}
An advertisement $\texttt{ad}=\{x_1,...,x_\ell\}, x_i \in \bin$ is a binary vector of length $\ell$.  Each index in the vector represents a particular (implicit) quality the media for that advertisement could encode. When a particular index is 1, that means the feature is present in the media. Importantly, we use this formalism to describe a piece of media directly, rather than allowing an advertiser to present
media and then
choose
a binary vector associated with that media; in this way, we assume that it is \emph{impossible} for an advertiser to lie about the features of an ad. 
\end{definition}

\begin{definition}[Audience] 
\label{defn:audience}
An $\texttt{audience}=\{x_1,...,x_\ell\}, x_i \in \bin$ is
also
a binary vector
of
length $\ell$.  Each index in this vector encodes an attribute that members of the audiences should have.  We assume that the meanings of
indices
for advertisements and audiences are consistent with one another---that is,
the $i^{\text{th}}$ element of each
encodes the same feature.
\end{definition}

\smallskip
\noindent
\textbf{Model overview.}  We give a high-level depiction of our model in \cref{fig:frameworkoverview}.  Namely, our model consists of five main ideal functionalities: $\idealsocietyfull$, $\idealuserfull$, $\idealtargetingfull,$ $\idealengagementfull,$ and $\idealmetricsfull.$   
In \cref{fig:society},
$\idealsocietyfull$ is responsible for sampling the features for each of the $n$ users in the system from some distribution $\mathcal{D}.$  Importantly, this means that the exact features for each user is hidden from the environment---although the distribution $\mathcal{D}$ may be known to the environment.
In \cref{fig:user},
$\idealuserfull$ is a subroutine that serves as the shared data infrastructure of the entire system, including holding each user's features and information about their interactions with the advertising system.  We note that $\idealuserfull$ might be implemented in a distributed manner, such that different elements of the data may be held by different real-world computational parties.
In \cref{fig:targeting,fig:engagement,fig:metrics},
$\idealtargetingfull$,
$\idealengagementfull$,
and $\idealmetricsfull$ 
make up the core of the advertising ecosystem.
Concretely,
$\idealtargetingfull$
is
responsible for choosing advertisements to deliver to users, $\idealengagementfull$
is
responsible for determining the websites that a user visits and how a user will interact with advertisements on those websites, and $\idealmetricsfull$
is
responsible for attributing conversion events and reporting on the performance of advertisements.

\begin{figure}[t]
\begin{center}
	\begin{tcolorbox}[enhanced,sharp corners,colback=white,boxrule=0.3mm,left=1mm]
		\vspace{0.5\baselineskip}
		Ideal Functionality $\idealengagementfull$

		\vspace{0.5\baselineskip}
		\hrule 
		\vspace{0.5\baselineskip}

        \idealengagement is parameterized by  \browsingfunction that selects a site for a user to visit and  \engagementfunction that determines how a user will interact with an advertisement. 
        
        \smallskip
        \textbf{Browsing:} Upon receiving an \texttt{browsing} message for $User_i$ from \Adv:
        \begin{enumerate}[noitemsep]
            \item Send a \texttt{browsing} message to \idealuser for $User_i$.
            \item Upon receiving a response from \idealuser for $User_i$ with \texttt{features}, and full \texttt{browsing-history}, generate $\texttt{site} \leftarrow \browsingfunction (\texttt{features}, \texttt{browsing-history}).$
            \item Send $site$ to \idealuser for $User_i$.
            \item Upon receiving \texttt{ok} from \idealuser, send \texttt{ok} message to \Adv.
        \end{enumerate}

        \smallskip
        \textbf{Ad Engagement:} Upon receiving an \texttt{engagement} message for $User_i$ from \Adv:
        \begin{enumerate}[noitemsep,nolistsep]
            \item Send an \texttt{engagement} message to \idealuser for $User_i$.
            \item Upon receiving a response from $\idealuser$ with $User_i$'s \texttt{features} and a tuple $( \texttt{site},\texttt{ad}),$ generate $\texttt{conversion} \leftarrow \engagementfunction(\texttt{features}, \texttt{site}, \texttt{ad})$. Note that $\texttt{conversion}$ may be \texttt{None}.
            \item Respond to \idealuser with $User_i$, $\texttt{ad}$, and $\texttt{conversion}$. 
            \item Upon receiving \texttt{ok} or \texttt{fail} from \idealuser, send \texttt{ok} message to \Adv.
        \end{enumerate}

	\end{tcolorbox}	
\end{center}
\caption{Engagement functionality $\idealengagement$}
\label{fig:engagement}
\end{figure}

\smallskip
\noindent
\textbf{Model flow details.} Next,
we illustrate
how
these functionalities work
and provide
a detailed description of the way data flows through the system.  We note that some of our functionalities also allow for interactions to occur in a different order.

\begin{enumerate}[left=0pt,label=(\arabic*),font=\bfseries,itemsep=0pt,wide]
    \item \textbf{Populating user features:} The features associated with each user are set up on demand.  Specifically, $\idealsocietyfull$ is set up with a total number of individuals
    $n$ that
    it will create
    and a distribution from which each individual's features will be sampled.  The environment does not need to explicitly initiate this sampling process, as $\idealsocietyfull$ will perform this ``just in time'' whenever $\idealuser$ encounters a user with no recorded features.

    \item \textbf{Registering Ad Campaigns:} When an advertiser wants to send an advertisement, they begin by sending a \textbf{Register Campaign} message to $\idealtargetingfull$ (\cref{fig:targeting})
    that specifies
    the explicit target \texttt{audience} to which they want to advertisements to be shown as well as the features of the advertisements $\{\texttt{ad}_1, \ldots, \texttt{ad}_k\}$ in the campaign (e.g., embedded within the visual media).  We emphasize that the modeling is done such that the advertiser cannot ``lie'' about the semantic content of the advertisement---the feature vector \textit{is} the advertisement. 

    \item \textbf{Initiating browsing:} The environment prompts the user to browse a website by calling the $\textbf{Browsing}$ interface of $\idealengagementfull.$  Note that the environment does not know the specific features of any given user, so we don't have the environment specify the website directly.  Rather, we use $\browsingfunction$ to choose the website that the user visits, possibly based on the user's features.  Specifically, 
    $\idealengagement^{\browsingfunction,\engagementfunction}$ requests $User_i$'s features from $\idealuser$ and obtains the \texttt{site} using $\browsingfunction$, which is defined over the features of the user and their previous browsing history.
    Then,
    $\idealengagementfull$
    informs $\idealuser$ that $User_i$ has visited $\texttt{site}$.

    \item \textbf{Advertisement Targeting and Delivery:} 
    To model the delivery of an advertisement to a user that has been prompted to visit a website, the environment uses the \textbf{Target Ad} interface of $\idealtargetingfull.$  This triggers a \texttt{target} message to $\idealuserfull$ in order to obtain the necessary information about the user and the context (i.e., \texttt{site}) in which the ad will be displayed.
    Next,
    $\idealtargetingfull$
    uses $\rho$ and $\targetingfunction$ to select the ad that will be shown to the user.  Note that the input to $\targetingfunction$ should operate over the the \texttt{audience} associated with the advertisements. The chosen advertisement is then sent to $\idealuserfull$ to be stored.

    \item \textbf{User Engagement:} After the user has viewed the advertisement (i.e., $\idealuserfull$ holds a tuple containing a \texttt{site} and an \texttt{ad}), the environment triggers possible user engagement using the \textbf{Ad engagement} interface of $\idealengagementfull.$
    In response,
    $\idealengagementfull$ retrieves the necessary information from $\idealuserfull$ and, using the parameterizing function $\engagementfunction,$ determines if the user turns generates a conversion on that impression.   Note that the input to $\engagementfunction$ is the features associated with the advertising media \texttt{ad} (as the user actually sees the \emph{media}, not the target \texttt{audience}).
    The results of this determination are then stored back in $\idealuserfull$.

    \item \textbf{Attribution:}  Before any metrics information can be provided, $\idealmetricsfull$ must first attribute each conversion event to at least one impression.  The environment prompts this through the \textbf{Attribute} interface of $\idealmetricsfull.$  When invoked in this way, $\idealmetricsfull$ calls to $\idealuserfull$ and retrieves the user's conversion history.
    Then,
    $\idealmetricsfull$
    updates the ``scores'' of each ad based on the output $\attributionfunction.$  It is easiest to think of this
    step
    as attributing the full ``credit'' for the conversion to the last impression.

    \item \textbf{Report Creation:} Finally, the environment (as the advertiser)  requests a report on the performance of its campaign.  To do this, the environment invokes the \textbf{Generate Report} interface of $\idealmetricsfull$, specifying a campaign (i.e., a set of advertisements).  These are then transformed into a report by the $\reportcreationfunction$ function, which is also responsible for adding noise or any other
    privacy protection
    mechanism.
\end{enumerate}

\begin{figure}[t]
	\ifdefined\cryptolayout
	{\footnotesize
	\fi
\begin{center}
	\begin{tcolorbox}[enhanced,sharp corners,colback=white,boxrule=0.3mm,left=1mm]
		\vspace{0.5\baselineskip}
		Ideal Functionality $\idealmetrics^{\attributionfunction,\reportcreationfunction}$

		\vspace{0.5\baselineskip}
		\hrule 
		\vspace{0.5\baselineskip}

        \idealmetrics is parameterized by an attribution function \attributionfunction and report generation function \reportcreationfunction.  Additionally, \idealmetrics maintains an updatable map \texttt{ad-scores}.

        \smallskip
        \textbf{Attribute:} Upon receiving an \texttt{attribute} message from \Adv for $User_i$: 
        \begin{enumerate}[noitemsep]
            \item Send a \texttt{attribute} message to \idealuser for $User_i$.
            \item Upon receiving a response from \idealuser for $User_i$ with list \texttt{browsing-history}, run $[(\texttt{ad}_1,\texttt{score}_1), (\texttt{ad}_2,\texttt{score}_2), ...] \gets \attributionfunction(\texttt{browsing-history}).$
            \item For each $\texttt{ad}_i$ in the output, add $\texttt{score}_i$ to the entry for $\texttt{ad}_i$ in \texttt{ad-scores}.
            \item Send an \texttt{ok} message to \Adv.
        \end{enumerate}

        \smallskip
        \textbf{Generate Report:} Upon receiving a \texttt{Report} message from \Adv for a $\texttt{campaign}:(\texttt{audience}, \{\texttt{ad}_1, \ldots, \texttt{ad}_k\}),$
        \begin{enumerate}[noitemsep,nolistsep]
            \item Let $\texttt{ad-scores}|_{\texttt{campaign}}$ be a map that is a subset of $\texttt{ad-scores}$ such that 
            \begin{multline*}
            \texttt{ad-scores}|_{\texttt{campaign}} = \\ \left\{ (\texttt{ad}_i,\texttt{score}_i) \in \texttt{ad-scores} ~|~ \texttt{ad}_i \in \{\texttt{ad}_1, \ldots, \texttt{ad}_k\} \right\}.
            \end{multline*}

            \item Generate $\texttt{report} \gets \reportcreationfunction(\texttt{ad-scores}|_{\texttt{campaign}})$.
            \item Respond to \Adv with a message containing \texttt{report}.
        \end{enumerate}

	\end{tcolorbox}	
\end{center}
\ifdefined\cryptolayout
}
\vspace{-2em}
\fi
\caption{Metrics functionality $\idealmetrics$}
\label{fig:metrics}
\ifdefined\cryptolayout
\vspace{-2em}
\fi
\end{figure}

\section{Inherent Tension Between Privacy and Usefulness}
\label{sec:privacyvsusefulness}

Now that we have introduced our
abstract
modeling of the advertising ecosystem,
in this section we
formalize two key concepts: (i) what does it mean for this ecosystem to be ``useful'' (what is the minimal functionality we need from our parameterizing functions) and (ii) what does it mean to add privacy to this ecosystem.
We use this formalism to show, analytically and empirically, that privacy and utility are inherently in tension.

\subsection{Defining Utility}
\label{sec:usefulfunctions}

\noindent
\textbf{\texttt{active-ads}.} In order to make the notation more direct, we define a set \texttt{active-ads} that represents the set of advertisements from which targeting may choose.  Specifically, for any \texttt{active-campaigns}, let \texttt{active-ads} be defined as follows:
    \begin{multline*}
        \texttt{active-ads} = \{ (\texttt{audience}, \texttt{ad} ) ~|~ \\  \{\texttt{audience}, \texttt{ad-set}\} \in \texttt{active-campaigns}, \texttt{ad} \in  \texttt{ad-set} \}.
    \end{multline*}

\noindent
\textbf{Measuring closeness.} 
We also require a concept of \textit{relevancy} to capture the idea that behavioral advertising is intended to show users ads that are relevant, or closely matched, to their interests and demographics.
We represent this with a $close$ \emph{metric} that takes as input two (binary) feature vectors and outputs a score that increases as the distance between the inputs shrinks.

\smallskip

\label{sec:formal-utility}
\noindent
\textbf{Targeting.} We begin with our utility function for targeting.  Specifically, a useful targeting system should be one that delivers
ads
that are more relevant to people with higher probability.  We formalize this notion by saying that the probability that an one advertisement is chosen over another is proportional to the difference in $close$ between the targeting audience and the user's features.\footnote{In practice, \textit{close} should also take in \texttt{site} as an input. However, since this context is, in theory, just a coarse-grained view into a user's features, we ignore it in order to simplify our analysis.}

\begin{definition}[Targeting Utility]
\label{defn:utility:targeting}
    A targeting function $\targetingfunction$ is $\alpha$-\emph{useful} with respect to a distance measurement $close$ and filter function $\rho$ if, given inputs 
    \texttt{active-campaigns}, \texttt{features}, and \texttt{site}, for all $(\texttt{audience}_1, \texttt{ad}_1), (\texttt{audience}_2, \texttt{ad}_2) \in \texttt{active-ads},$ if
    \begin{multline*}
    close(\texttt{audience}_1, \texttt{features}) - \\ close(\texttt{audience}_2, \texttt{features}) = \close, \text{ then, } \\ 
        \Pr\left[\texttt{ad}_1 \leftarrow \targetingfunction(\texttt{active-campaigns}, \rho(\texttt{features}),  \texttt{site})\right]  - \\ \Pr\left[\texttt{ad}_2 \leftarrow \targetingfunction(\texttt{active-campaigns}, \rho(\texttt{features}), \texttt{site})\right]  \geq \alpha \cdot \close.
    \end{multline*}

\end{definition}

\noindent
\textbf{Engagement.} A foundational assumption of advertising is that individuals are more likely to engage with advertisements that are
more
``like them.'' We 
formalize this idea
using a
closeness metric, similar to the one in Definition~\ref{defn:utility:targeting} for
targeting utility.
\begin{definition}[Engagement Utility] \label{defn:utility:engagement} We say that an engagement function \engagementfunction is $\alpha$-\emph{useful} with respect to a distance measurement $close$ if for
any set
of user features $\texttt{features}$,
website $\texttt{site}$, pair of advertisements $(\texttt{ad}_1, \texttt{ad}_2)$,
and non-None conversion
event
$\texttt{conversion}$:
\begin{multline*}
\text{if }close(\texttt{ad}_1,\texttt{features}) - close(\texttt{ad}_2, \texttt{features}) = \close,
\\
\end{multline*}
\vspace{-0.45in}
\begin{multline*}
\text{then }\Pr\left[\texttt{conversion} \leftarrow \engagementfunction(\texttt{features}, \texttt{site}, \texttt{ad}_1) \right] \\ - \Pr\left[\texttt{conversion} \leftarrow \engagementfunction(\texttt{features}, \texttt{site}, \texttt{ad}_2) \right] \geq \alpha \cdot \close.
\end{multline*}
\end{definition}

\noindent
\textbf{Attribution.} Attribution is considered useful if it is more likely to attribute a conversion to the impression that generated it than an unrelated impression. For the purposes of our analysis, we model utility of \idealmetricsfull relative to the ground truth as generated by \idealengagementfull. One of the limitations of our model is that \idealengagementfull does not model cases where many impressions contribute to a single conversion event. So, while our attribution functionality is generic and can handle multi-touch attribution, the ground truth for this analysis is that a \textit{single} ad is responsible for each conversion.

\begin{definition}[Attribution Utility] \label{defn:utility:attribution} 
An attribution function $\attributionfunction$ is $\alpha$-\emph{useful} with respect to an engagement function $\engagementfunction$ if for any feature vector \texttt{features}, any pair of advertisements $(\texttt{ad}_1,\texttt{ad}_2)$ and associated websites $\texttt{site}$, and any non-None conversion event \texttt{conversion}:
\begin{multline*}
\text{if }\Pr\left[\texttt{conversion} \leftarrow \engagementfunction(\texttt{features}, \texttt{site}, \texttt{ad}_1) \right] - \\ \Pr\left[\texttt{conversion} \leftarrow \engagementfunction(\texttt{features}, \texttt{site}, \texttt{ad}_2) \right]=\close,
\end{multline*}
\vspace*{-0.25in}
\begin{multline*}
  \text{then }\Pr\left[ scores[\texttt{ad}_1] > scores[\texttt{ad}_2] \right. ~|~ \\ \left. scores \leftarrow \attributionfunction(\texttt{conversion}, \texttt{browsing-history})  \right] \geq \alpha \cdot \close
 \end{multline*}
if \texttt{browsing-history} contains both $\texttt{ad}_1$ and $ \texttt{ad}_2$.

\medskip

\noindent
\textbf{Metrics.} Metrics is considered useful if it permits statistical tests to be conducted on the results. That is, if some test, such as an A/B test, could be conducted on the raw attribution data, it should still be possible to conduct this test on the aggregated and possibly noisy version of this data output by metrics. That is, the utility of metrics is defined based on what the advertiser intended to do with the attribution data.

\begin{definition}[Metrics Utility Preserving]
\label{defn:utility:metrics}
    For all $h$, we say that a randomized metrics report generation function $\reportcreationfunction:\mathcal{D}^h \rightarrow \mathcal{D}^h$ is \emph{$\alpha$-utility-preserving} with respect to a (possibly randomized) processing function $f_s:\mathcal{D}^h \rightarrow \bin$ if
    for all
    $ 
    \hat{d}=\{d_1,...,d_h\} \in \mathcal{D}^h,$
    $$ \lvert \Pr[f_s(\hat{d})=1] - \Pr[f_s(\{\reportcreationfunction(
    \hat{d}
    )\})=1] \rvert <\alpha,$$
    where the probabilities are over the randomness of $ \reportcreationfunction$ and $f_s$.\footnote{In practice, this definition requires that a statistical test applied to the output of the report generation function will still provide the same result as the test on the raw data, albeit with an error rate of $\alpha$. This can also be thought of as requiring the same result of a t-test with a worse p-value.}
\end{definition}

\end{definition}

\subsection{Formal Statement}\label{sec:proof}\label{sec:proof:mainbody}

In this section, we state and prove our formal result: there is an innate tension between preventing leakage and preserving utility in an advertising ecosystem.  

\begin{theorem}
\label{thm:utilityimpliesleakage}
Any ads ecosystem composed of instantiations of $\idealtargetingfull,$ $\idealengagementfull,$ and $\idealmetricsfull$ that are \emph{useful} (as defined by \cref{defn:utility:targeting,defn:utility:engagement,defn:utility:attribution,defn:utility:metrics}, with the additional restriction that any non-trivial implementation of $\idealmetricsfull$ must use differential privacy) for a given ad campaign will not satisfy attribute privacy for some attribute of that campaign's audience.\footnote{This is not to say that the theorem cannot hold for other instantiations of $\idealmetricsfull$, however, all existing private metrics proposals make use of differential privacy so we make our proof in this setting as well. We additionally require $\epsilon<1$.}
\end{theorem}

We prove this theorem by showing that anything that could be learned by an advertiser in a non-private advertising system could similarly be learned by advertiser in a private advertising system.  Proving this statement formally requires defining a game-based privacy definition \emph{on top} of our UC modeling.
To that end, we define the following random variable.

\begin{definition}\label{def:exec}
    Let $\mathsf{EXEC}^{\targetingfunction, \rho, \browsingfunction,\engagementfunction,\attributionfunction,\reportcreationfunction,n,\mathcal{D}}_{\Adv}$ be a random variable denoting the output distribution of an environment $\Adv$ when interacting with the ideal functionalities $\idealtargetingfull$, $\idealuser$, $\idealengagementfull$, $\idealmetricsfull$ and $\idealsocietyfull$ (connected as shown in \cref{fig:frameworkoverview}) in an instance of the UC experiment.
\end{definition}

Typically in a game-based definition, we have a \emph{challenger} that sets up the parameters of the game (sampling randomness as needed) and an \emph{adversary} that is required to guess some function of the \emph{challenger}'s randomness. 

\begin{definition}[Distinguishing] 
\label{defn:privadsystem}
 We say that an adversary $\mathcal{A}=(\mathcal{A}_0,\Adv,\mathcal{A}_1)$ succeeds in distinguishing with probability $p$ with respect to a distribution $\mathcal{D}_0$,  processing function $f_s\mathcal{D}^h\rightarrow\bin$, and an advertising system defined by the parameters $(\targetingfunction,\rho,\browsingfunction,\engagementfunction,\attributionfunction,\reportcreationfunction,n)$ if:
$$p = 2 \cdot \Pr [\mathcal{A}_1(f_s(\mathsf{EXEC}^{\targetingfunction,\rho,\browsingfunction,\engagementfunction,\attributionfunction,\reportcreationfunction,n,{\mathcal{D}_b}}_{\Adv(\mathcal{D}_1,aux)}),aux) = b ] - 1,$$
where $(\mathcal{D}_1,aux) \gets \mathcal{A}_0(1^\lambda, \Dzero)$ and $b\sample\bin$, and the probability is taken over the random choices of $\adv$, $b$, and the execution.
We define the adversary's advantage $\advantage{\targetingfunction,\rho,\browsingfunction,\engagementfunction,\attributionfunction,\reportcreationfunction, n}{\mathcal{A},\mathcal{D}_0,f_s} \coloneqq p$.
\end{definition}
In this definition,
$\mathcal{D}_0$ should be thought of as the \textit{ground truth} distribution of features across people in society, whereas $\mathcal{D}_1$ represents the advertiser's \textit{prior knowledge} about how people's features are distributed.
Hence, the distance between these distributions corresponds to the precision of information gained by distinguishing. When $\mathcal{D}_1$ is close to $\mathcal{D}_0$, distinguishing will be more challenging, but
the advertiser
can
learn finer-grained information \cite{canonne2020survey}.

With this notion in hand, we can now state that whenever there is an adversary that can succeed at distinguishing within non-private advertising systems, then there exists an adversary that can succeed in any private version of that system that preserves utility.  
More precisely,
a useful $\idealmetricsfull$ requires that the campaign size $n$ was large enough to still obtain a useful result from the output of $\reportcreationfunction^{\epsilon}$. Here,
$n$ plays the role
of \textit{sample complexity} from distribution testing, which is the number of samples necessary to distinguish between two distributions. Thus, our approach here is to show that with a sufficiently-sized campaign in the private setting, an adversary can learn the same information as in the non-private setting. Borrowing from the conventions
used in
distribution testing, we focus on the goal of distinguishing with advantage $\frac{2}{3}$ in the non-private setting.

Specifically, we prove the following 
two lemmas.

\begin{lemma}
\label{thm:distinguishing}
Let $\mathcal{A}=(\mathcal{A}_0,\Adv,\mathcal{A}_1)$ be an adversary.
Consider any two ad ecosystems:
\begin{itemize}[--, left=6pt]
    \item A non-private ecosystem with a targeting function $f_t$ that is $\alpha_t$-useful with respect to a $close$ metric and the identity function $I$ as the lens, an engagement function $f_e$ that is $\alpha_e$-useful with respect to $close$, an attribution function $f_a$ that is $\alpha_a$-useful with respect to $f_e$, and that uses the identity function $I$ for reporting.
    \item A private ecosystem with a (possibly different) targeting function $f_t'$ that is $\alpha_t'$-useful with respect to the same $close$ metric and filtering lens $\rho'$, and with a reporting function $f^{\epsilon}_r$, where $\epsilon<1$ that is $\alpha_r$-utility-preserving with respect to a processing function $f_s: \mathcal{D}^h\rightarrow\bin$.
\end{itemize}
For any distribution $\mathcal{D}_0$ over $\bin^\ell$, for $(\mathcal{D}_1,aux) \gets \mathcal{A}_0(1^\lambda,\Dzero)$, where $\mathcal{D}_1$ has the same support as $\Dzero$ and both are over the domain $\mathcal{X}$, and for any collection of \texttt{active-ads}: if
$\advantage{\targetingfunction,I,\browsingfunction,\engagementfunction,\attributionfunction,I, n}{\mathcal{A},\mathcal{D}_0,f_s} \geq \frac{2}{3}$,
then
$\advantage{\targetingfunction',\rho',\browsingfunction,\engagementfunction,\attributionfunction,\reportcreationfunction^\epsilon, n'}{\mathcal{A},\mathcal{D}_0,f_s} \geq \frac{8}{15}$, where $n' < \frac{100n}{\epsilon} \cdot \frac{1+\alpha_tK}{1+\alpha_t'K}$. Here, $K$ is a computable term that depends only on $\mathcal{X}$, \texttt{active-ads}, $\Dzero$, $\Done$ and $\frac{100n}{\epsilon} \cdot \frac{1+\alpha_tK}{1+\alpha_t'K} < \frac{100n}{\epsilon} \cdot \frac{\alpha_t}{\alpha_t'}$.
\end{lemma}

\begin{lemma}
\label{thm:distinguishingimpliesnoprivacy}
For any ads ecosystem where there exists an adversary with distinguishing advantage $\advantage{\targetingfunction,\rho,\browsingfunction,\engagementfunction,\attributionfunction,\reportcreationfunction, n}{\mathcal{A},\mathcal{D}_0,f_s} > 0$,
this ads ecosystem will not satisfy attribute privacy.
\end{lemma}

\medskip
\noindent
\textbf{Proof Sketch.} We formally prove \cref{thm:utilityimpliesleakage} in \cref{app:theorem-proof}. It follows immediately from \cref{thm:distinguishing,thm:distinguishingimpliesnoprivacy}.

To prove \cref{thm:distinguishing}, we show that utility implies the ability to distinguish the underlying distribution used by the ads ecosystem.
Namely, any successful distinguisher for a given advertising campaign in a non-private ads ecosystem can be used to construct a distinguisher for the same campaign run in a useful, private ads ecosystem, albeit with a larger campaign size.
The main idea behind our proof of \cref{thm:distinguishing} is that, while the private version of an ads ecosystem may have less utility than a non-private version, as long as it preserves \textit{some} utility, then it is possible
to amplify this signal to
match the utility
of the non-private version.
The ``cost'' of
amplification is in
increasing the size of the campaign $n$,
which
provides the adversary with more samples to use in its distinguishing. We can leverage distribution testing techniques \cite{bar2002complexity,canonne2019structure,cai2017priv} to find a bound on this new campaign size $n'$.\footnote{In practice, this demonstrates the disparate impact of differential privacy between large advertisers with huge campaigns versus smaller independent advertisers.}

The proof of \cref{thm:distinguishingimpliesnoprivacy} follows from the definition of attribute privacy (\cref{defn:datasetattributeprivacy}) and the ability of our distinguisher to identify the underlying distribution used by the private ads ecosystem. Specifically, if attribute privacy were achieved for all parameters governing the distribution of users in $\idealsocietyfull$, then by definition the summary statistic output by $\idealmetricsfull$ should be independent of changes to this distribution. However, were this the case, then the output of $\idealmetricsfull$ would be independent of the choice of $D_0$ or $D_1$ and no successful distinguisher could exist. Since \cref{thm:distinguishingimpliesnoprivacy} assumes the opposite, there must exist \textit{some} parameter of the underlying distribution for which attribute privacy is \textit{not} preserved.
We explore the idea that not every attribute may require such protection, i.e., that some inferences may be acceptable, in \cref{sec:redefining-privacy}. 

\subsection{Empirical Sample Complexity}
\label{sec:empirical-distinguishing}
To provide some intuition and empirical data for the concrete sample complexity increase that we showed theoretically in \cref{thm:distinguishing}, we implemented our ideal advertising functionalities in Python\footnote{\url{https://github.com/kylehogan/idealAdsFunctionalities}}
 and ran the distinguishing game from \cref{sec:proof:mainbody} for concrete realizations of our parameterizing functions and $\alpha$-utility parameters.

Recall that
$\alpha$-utility for targeting is an indication of how tightly the parameterizing function respects its definition of $close()$, or how \textit{accurate} targeting is able to be. 
The expectation is that private advertising ecosystems will likely have less of an ability to find the true closest ad for a user---whether through using less user data or less precise data---and we handle this challenge by decreasing the $\alpha$ parameter for targeting from the non-private version.
Relatedly,
$\alpha$-utility for engagement
represents how likely a user is to click on an ad at all; this is entirely
about user behavior,
so it
does not vary between
the
private and non-private ad ecosystems. 
Our metrics parameterizing function is instantiated using differential privacy, as this is the most common method currently being proposed.

\begin{figure}[t]
\includegraphics[width=0.45\textwidth]{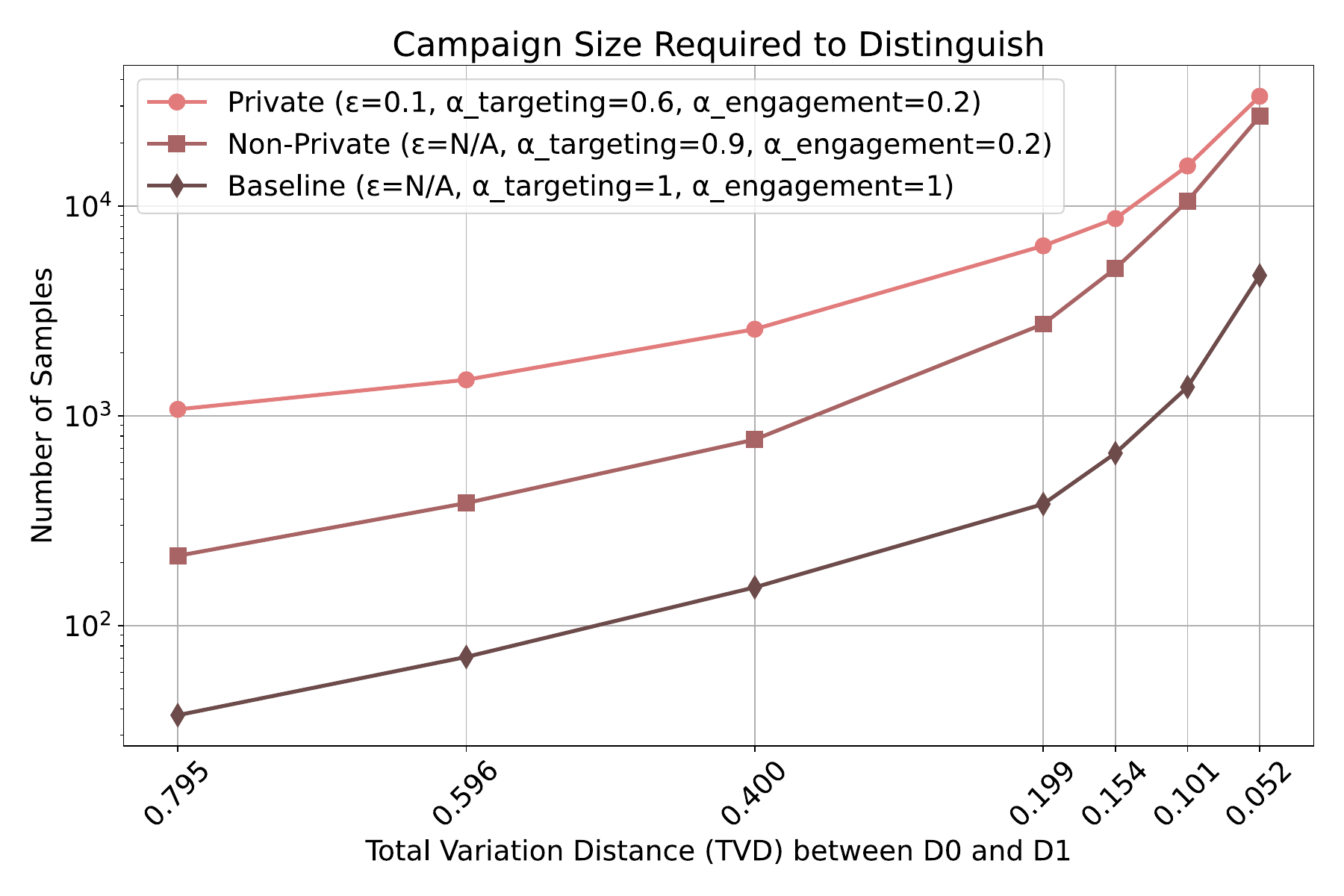}
\caption{Impact of privatization on sample complexity.}\label{fig:empirical-distinguishing}
\end{figure}

To run our empirical distinguishing game, we fix a campaign with two ads ($ad_A$ and $ad_B$) differing in a single bit $b_{test}$, and a distribution $\mathcal{D}_0$ representing the ``ground truth'' distribution of users. 
We then create an alternate distribution $\mathcal{D}_1$ based off the same covariance matrix as $\mathcal{D}_0$, varying the marginal probability of $b_{test}$ in $\mathcal{D}_1$ to gradually increase the total variation distance between $\mathcal{D}_0$ and $\mathcal{D}_1$. 
We plot the sample complexity required to distinguish $\mathcal{D}_1$ from $\mathcal{D}_0$ at a p-value of 0.05 using the uniformly most powerful tests from \citet{Awan_Slavkovic_2020} for a private ads ecosystem in \cref{fig:empirical-distinguishing}. Then, using a standard binomial test to distinguish, we plot a non-private version of the same ecosystem as well as a baseline to demonstrate the increase in sample complexity.

The `non-private' line still shows a substantial increase over the baseline due to the loss of accuracy from targeting and the drop in user engagement, while the `private' line indicates the impact of further reducing targeting accuracy and, more importantly, introducing noise from differential privacy. 
The private and non-private lines begin to converge at higher sample complexity due to the lower relative impact of differential privacy for larger sample set sizes.
We provide plots on the individual impacts of $\alpha$-targeting, $\alpha$-engagement, and $\epsilon$-differential privacy in \cref{app:empirical-plots}.

\section{Redefining Privacy for Advertising}
\label{sec:redefining-privacy}

Our results are a clear indication that the path forward for private advertising requires a careful re-imagining of what privacy \emph{should} mean.  That is, if we accept that some leakage is a necessary part of the advertising ecosystem, then how should advertising systems reason about the risks posed by this leakage? To start this process, we begin by introducing
the perspectives of three stakeholder groups
that inform this future: the people receiving targeted advertisements, advertising networks, and regulatory bodies. 
By identifying commonalities across these viewpoints---and gaps between existing policies and end-users needs---we hope a path forward can emerge. 

In \cref{sec:defining-sensitive},
we employ the contextual integrity framework \cite{nissenbaum2004privacy} to help delimit when leakage may be (in)appropriate. We contrast under what conditions users, ad tech, and regulatory bodies consider different categories of data to be sensitive\footnote{In an advertising context specifically, as opposed to \textit{generally} sensitive.}, and thus deserving of stronger protections. We then outline how existing enforcement mechanisms tailored to data sensitivity fail, despite laws and well-aligned ad tech policies. The principles underlying existing private metrics proposals offer technical solutions to some of these gaps, but
they
ultimately fall short of meeting user's privacy expectations.
We argue that future proposals for
privacy-preserving
metrics must be \textit{targeting-aware} in order to
(1)
discern between the implications of different information leakages
and
(2)
understand that risks associated with leakage are context dependent.
In doing so,
private metrics systems of the future
can
apply alternate notions of privacy, such as attribute privacy \cite{zhang2022attribute}, that can
protect
sensitive features about advertising audiences as a whole.

\subsection{\textit{Sensitivity} in Advertising}
\label{sec:defining-sensitive}

Users, advertising technology companies, and  regulators all agree that some types of information about individuals are \textit{sensitive} and inappropriate to use in an advertising context. 
In this section, we use the framework of contextual integrity \cite{nissenbaum2004privacy} to interrogate the conditions under which transmitting metrics from the ad network to the advertiser are appropriate.
Explicitly, contextual integrity considers the flow of specific information types about a subject between a sender and recipient via a transmission principle and determines whether this flow is \textit{appropriate}.
In the setting of this paper, ad networks (source) transmit various types of data (ad targeting features) about the users (subject) in ad audiences to advertisers (recipient) via metrics reports (transmission principle) about ad delivery and engagement. 
Whether this process is perceived as appropriate largely depends on what type of information was used in targeting, which is the focus of \cref{fig:sensitive-data}.

\begin{figure}[t]
    \setlength{\tabcolsep}{4.5pt}
    \renewcommand{\arraystretch}{1.2}
    \centering
    \begin{tabular}{|c||c|c|c|}
    \hline
    \multicolumn{4}{|c|}{\textbf{Appropriateness of Targeting on Feature}} \\
    \hline
        \multirow{3}{*}{Sensitive Data} & \multicolumn{3}{c|}{As perceived by...} \\
        & People & Ad Tech & Law \\ 
        & \scriptsize{\cite{schoenebeck2023sensitive,leon2015privacy,hanson2020taking,192371,wei2020twitter}} & \scriptsize{\cite{googleadpolicy}}& \scriptsize{EU\cite{dsa},US\cite{fha}}\\\hline
         health & \faBan & \faBan &  \faBan \\ \hline
        relationship & \faExclamationTriangle & \faExclamationTriangle & \faCheck \\ \hline
         political beliefs & \faBan & \faBan & \faBan\\ \hline
         sexuality & \faBan & \faBan & \faBan \\ \hline
         gender & \faCheck & \faBalanceScale & \faBalanceScale\\ \hline
         location & \faExclamationTriangle & \faBalanceScale &  \faBalanceScale\\ \hline
         age & \faCheck & \faBalanceScale & \faBalanceScale \\\hline
    \end{tabular}
    \caption{ \faCheck~ is used to indicate that targeting on this feature is acceptable/permitted, \faBan~ indicates that targeting is \textit{always} considered unacceptable/prohibited, \faBalanceScale~ indicates that targeting is permitted unless illegal (in the case of tech policy) or discriminatory (in the case of regulations), and \faExclamationTriangle~ indicates that targeting is \textit{conditionally} acceptable.}
    \label{fig:sensitive-data}
\end{figure}

Contextual integrity allows us to bring nuance to our data categorizations. It isn't that advertisements can \textit{never} feature health or sexuality information---in fact, it is often actively beneficial to promote awareness of mental health support options or to advertise events at LGBT organizations.
Thus, our focus is specifically on data sensitivity in the context using these data types (implicitly or explicitly) to \textit{target advertisements} and \textit{conduct market research}, not simply displaying ads.\footnote{We again note that contextual advertising is a type of targeting and it is potentially still inappropriate to conduct market research over ads that were, for example, shown only on LGBT-focused webpages.}

The strongest ad tech policies \cite{googleadpolicy} and modern, advertising-specific regulations like the Digital Services Act (DSA) \cite{dsa} both align quite closely with user preferences.  
Unfortunately, it has proved difficult to enforceably put these policies into practice.

\smallskip
\noindent
\textbf{Existing policies are sophisticated and nuanced.}
While some types of personal data (like health data) are \textit{always} considered sensitive in the context of ad targeting~\cite{dsa,googleadpolicy,hanson2020taking}, other features can be more subtle.  
For example, while relationship status is used to target both ads for dating services and those for divorce lawyers, people are far less comfortable with the latter than the former, despite the same data being used in both cases~\cite{schoenebeck2023sensitive}.
This sentiment is captured by ad tech policy, which prohibits targeting based on ``personal hardship''---such as divorce---or advertisements that ``impose negativity'' (e.g., body shaming).
Similarly, targeting on the basis of features that people generally find acceptable, like age or gender, is illegal when the impact of that advertising results in \emph{discriminatory} systems.
As a result, regulations such as the Fair Housing Act (FHA)~\cite{fha} have been used to prohibit use of characteristics like age, race, and gender for all housing or employment advertisements in the United States \cite{fhadecision} and to enact changes to the targeting algorithm in the same vein~\cite{Timmaraju_2023}.

\smallskip
\noindent
\textbf{Enforcement problems stymie policies' promise.} 
Advertisement targeting is extremely opaque and largely built on machine learning models, making both technological and legal enforcement of these policies challenging~        \cite{calderonio2024fledging,277076,algotrithmwatchfb}.  The behavior of machine learning models is prohibitively difficult to interrogate, making
it challenging to prove
discrimination
\cite{wachter2020affinity,imana2021auditing,datta2018discrimination,speicher2018potential,203840,291188}. 
Despite some efforts on the part of ad networks to mitigate bias \cite{Timmaraju_2023}, it has been found that these targeting systems distribute advertisements in a discriminatory way \emph{even when it is not the intention of advertisers} \cite{291188}. 
It is also possible to intentionally circumvent protections using proxy features or ``lookalike audiences''
\cite{lookalikesgoogle,andreou2019measuring,wei2020twitter}.
Circumventing protections this way is, of course, against policy, but even the ad networks themselves have been caught using an opaquely-defined audience to illegally target ads to children \cite{googleignoresrules}. 

Often the opaque nature of targeting allows companies to avoid accountability with initial lawsuits struggling to prove discrimination \cite{opiotennione,vargas}. 
Moreover, it took until late 2023 for courts to recognize that ad networks, not only advertisers, are liable for the discriminatory targeting of ads \cite{liapes}. 
Successful litigation has often had to circumvent the root problem of the \textit{use} of sensitive data in targeted advertising to instead focus on how that data was \textit{collected}, relying on regulations for deceptive business practices \cite{ftctwitter} or even wiretapping \cite{linkedinsued,popa}.  This makes it burdensome for users to enforce their rights.
Finally, while advertising is global, regulations decidedly are not and this ultimately limits the ability of even the strongest regulations to protect the privacy of all users.

\subsection{Metrics is \emph{Sensitivity} Agnostic}
\label{sec:dp-metrics-insufficient}
If there
existed
meaningful enforcement of existing laws and ad targeting policies---and confidence that these strong laws and policies applied across the full advertising ecosystem---then perhaps we would not need to be as concerned about information leakage.
But
without such enforcement, there is a real risk that the information leaking from the system will directly concern sensitive data.
In this section,
we turn our attention to metrics in the hope that it can make up for the identified failures of targeting. 

\smallskip
\noindent
\textbf{Metrics is well-positioned to facilitate policy enforcement.}
Metrics does not face the same structural challenges that make aligning targeting systems and people's privacy preferences so difficult.
First,
the fraught (and legally tricky) decisions on which advertisements should be shown to which users have already been made.
Second,
the systems that collect and compute metrics are dramatically simpler and more transparent than those used to target advertisements. Thus, the metrics infrastructure could aid in identifying and documenting policy violations. 

Users also have significant agency that they can exert when it comes to metrics.  While users have no choice in the advertising networks to which they are subjected while browsing the internet, users' choice of browsers and devices are directly tied to the way their data is collected and processed within metrics.  In principle, this creates an opportunity for organizations to compete in order to make their metrics systems as well-aligned with user's privacy preferences as possible. Indeed, different groups of ad tech companies are currently working on competing proposals for privatizing metrics ~\cite{ppa,ara}. Importantly, these proposals are designed to be \emph{interoperable}, meaning that no matter which system was used to target an advertisement, a variety of organizations, each offering a different suite of privacy protections, are capable of producing equivalent metrics output.\footnote{User choice alone is likely insufficient to ensure that people's privacy is protected according to their preferences---default settings and other dark patterns are often successful in preventing users from exercising their ability to choose effectively \cite{bosch2016tales}.}

\smallskip
\noindent
\textbf{Current proposals fall short.}
We identify three reasons why current proposals for privacy-preserving metrics do not adequately enforce policy. First, their
technical underpinnings
rely on aggregation \cite{prio} and the injection of statistically-calibrated noise (i.e., differential privacy \cite{TCC:DMNS06}).  The result is an implicit understanding that privacy in advertising is about preserving the \emph{confidentiality} of individuals' features. As we observe in this work, however, some amount of leakage is inherent, and the leakage these systems permit is fundamentally de-contextualized, i.e., it is at odds with the understanding that \emph{not all types of data should be treated the same}, as demonstrated by \cref{fig:sensitive-data}.

Second,
current metrics proposals are unaware of the content and target audience of the advertisements whose performance they measure. Thus, an ad campaign promoting clothing is treated identically to an ad campaign promoting therapy, despite the difference in the sensitivity of the data likely used to target these advertisements.  Similarly, an ad campaign that uses gender to target clothing advertisements is indistinguishable from one that uses gender to target employment advertisements, despite the difference in how the law sees these campaigns.  This is intentional; existing metrics systems embraced data minimization within their design, and differentiating between advertisements or audience would require collating this data across multiple systems (i.e., from targeting systems to metrics systems).  While data minimization is generally the right approach for system design, in this case it has rendered metrics incapable of discerning between information leakages that people might consider harmful and innocuous.

Third,
existing metrics proposals
treat differential privacy as a privacy panacea, when,
in fact,
there are cases in which inference itself can be harmful (see \cref{sec:defining-privacy:nontrivial}).  Specifically, there are audience types, so called ``custom audiences,'' defined using personally identifiable information.  In these cases, the inference facilitated by differential privacy has qualitatively different risk as learning a feature of their audience also gives them confidence that \textit{individual members} of the audience possess this feature, contrary to the likely expectations of those audience members. 

\subsection{Closing the Gap}
While data sensitivity provides clear intuition for managing the risks of information leakage, the existing paradigms within which advertising systems are designed are insufficient to actualize this approach.  Within targeting, there has been significant policy work to set standards for the treatment of sensitive data, but there are structural barriers to enforcing these policies.  On the other hand, emerging metrics proposals are technologically sophisticated and relatively transparent, but are \textit{incapable} of enforcing normative privacy policies because they lack context on how ads were targeted.

In order to close this gap, we advocate for expanding the approaches that are being used to think about privacy when developing new targeting and metrics proposals.
There is tremendous, ongoing technical work integrating differential privacy into metrics computation in which researchers are leveraging cutting-edge privacy-enhancing technologies to significantly improve people's concrete privacy~\cite{ppa,tholoniat2024cookie}. These efforts, however, cannot be the sum total of the solution.  Specifically, future developments need to apply the same, policy-oriented analyses to metrics that are currently being applied to targeting. 
We advocate for the inclusion of group privacy notions, like attribute privacy, that explicitly account for privacy harms not covered by differential privacy.
Namely, as we introduced in \cref{sec:defining-privacy:nontrivial} and expanded on here, \textit{what} data is leaked can be just as, if not more, important than \textit{how much} information is revealed. This is especially true because advertisers can combine ``private'' metrics with already-known information about individual members of the  audience, such as their identities \cite{seeman2024between}.

However, applying attribute privacy to advertising metrics requires co-design across targeting and metrics protocols as, while targeting possesses the necessary information about data sensitivity, metrics does not. 
Distributional privacy notions naturally require information about the underlying distribution of users targeted by an advertisement which is not currently available to metrics protocols.
Making metrics systems \emph{targeting-aware} by giving it the audience information for its reports would allow for the use of definitions like attribute privacy and could, perhaps, even permit metrics protocols to monitor targeting for policy violations or discriminatory behavior. 
When targeting and metrics are run by different organizations, there may even be incentives to do this type of mutual monitoring.  While there will no-doubt be significant technical (and even legal) hurdles in implementing such a vision, the result would be better alignment between the privacy preferences of users and the privacy properties of advertising ecosystems.

\section{Conclusion}

In this work we have taken a step back to study what notions of privacy are possible within advertising.  We showed that any advertising system that is even minimally useful must also allow some amount of information leakage.  Taking this as a given, we identify the sensitivity of data as an important consideration when it comes to managing this leakage---a decision which has significant implications on how future privacy-preserving advertising proposals should be designed.

\begin{acks}
This research was supported by the DARPA SIEVE program under Agreement No.\ HR00112020021 and by the National Science Foundation under Grants No.\ 1955270, 2209194, 2217770, 2228610, 2230670, and 2330065.
\end{acks}

\bibliographystyle{ACM-Reference-Format}
\bibliography{extra_citations,abbrev3,crypto}
\appendix

\section{Proof of \cref{thm:distinguishing}}

In this appendix, we provide a formal proof of \cref{thm:distinguishing}.
Suppose $\mathcal{A}=(\mathcal{A}_0,\Adv,\mathcal{A}_1)$ is an adversary such that $\advantage{\targetingfunction,I,\browsingfunction,\engagementfunction,\attributionfunction,I, n}{\mathcal{A},\mathcal{D}_0,f_s} > \frac{2}{3}$ for a non-private ad ecosystem. We proceed via a sequence of games to show that this implies that $\mathcal{A}$ still has advantage $\advantage{\targetingfunction',\rho',\browsingfunction,\engagementfunction,\attributionfunction,\reportcreationfunction^\epsilon, n'}{\mathcal{A},\mathcal{D}_0,f_s} > \frac{8}{15}$ in the private case. That is: the adversary can continue to have a similar advantage in the private case, so long as the campaign size is appropriately amplified. 

\paragraph{Notation.}
In each game $\mathcal{G}_i$, we define $\advantage{i}{\adv}$ to be the advantage of adversary $\mathcal{A}^i=(\mathcal{A}^i_0,\Adv^i,\mathcal{A}^i_1)$ in distinguishing between executions of the specified games starting from input distributions $\mathcal{D}_0$ versus $\mathcal{D}_1$. For the first and last games, our objective is to distinguish between  distributions $\mathcal{D}_0$ and chosen distribution $\mathcal{D}_1$.
However, some of our intermediate games are defined with different types of outputs that correspond to partial execution of a hybrid between the non-private and private ad ecosystems. Between each game we will show that the adversary's advantage is preserved with the appropriate increase to the campaign size---which corresponds to the sample complexity of distinguishing between the distributions.

\paragraph{Overview.}
To summarize our approach, the goal of this proof
is to relate the adversary's advantage of distinguishing in the private case with their advantage of distinguishing in the non-private game in terms of the relative increase in the sample complexity required to distinguish.
Recall that, in our setting, the sample complexity is the campaign size.
The first game $\mathcal{G}_0$ represents distinguishing between the two distributions in the non-private setting, and the final game $\mathcal{G}_4$ is in the private setting.
Each game $\mathcal{G}_i$ makes a slight modification to the advertising ecosystem from the game before it $\mathcal{G}_{i-1}$, and throughout this sequence of hybrids we show that the adversary still has a non-trivial advantage in distinguishing given a certain increase in the number of samples available.
These games primarily leverage how the utility definitions require the two distributions to have certain properties to have been distinguishable in the non-private setting. We then leverage results and techniques from the distribution testing literature to derive the updated sample complexity and distinguishing advantage for each game.

\paragraph{Facts.}
To begin, we state a few facts relating Hellinger distance and sample complexity that we will reference throughout the proof:

\newtheorem{fact}{Fact}

\begin{fact}[Folklore, restated in \cite{canonne2019structure}]\label{fact:npbound}
The Hellinger distance characterizes the sample complexity $SC^{P,Q}=\Theta(1/H^2(P,Q))$ between two distributions $P,Q$.
\end{fact}

\begin{fact}[\cite{bar2002complexity}, Theorem 4.7]\label{fact:nplower}
If a distinguisher for two distributions $P,Q$ has error $\beta$, then the sample complexity for distingushing between these distributions is lower bounded as follows: $SC^{P,Q} > \frac{\textup{ln}(\frac{1}{4\beta})}{4\cdot H^2(P,Q)}$
\end{fact}

\begin{fact}[\cite{canonne2019structure},  Corollary 2.2]\label{fact:npupper}
The sample complexity for distinguishing between two distributions $P,Q$ is upper bounded as follows: $SC^{P,Q}  < \frac{1}{H^2(P,Q)}$.
\end{fact}

\begin{fact}[\cite{cai2017priv},  Theorem 2]\label{fact:private}
If the sample complexity $SC^{P,Q}$ for distinguishing two distributions $P,Q$ with success probability $\geq 1-\beta$ is $n$, then the sample complexity for distinguishing between these distributions $P,Q$ subject to $\epsilon$-differential privacy (where $\epsilon<1$) $SC_{\epsilon}^{P,Q}$ with success probability $\geq \frac{4}{5}(1-\beta)+\frac{1}{10}$ is bounded as follows: $n \leq SC_{\epsilon}^{P,Q} \leq \frac{10}{\epsilon\cdot H^2(P,Q)}=\frac{10n}{\epsilon}$.
\end{fact}

Recall that the Hellinger distance for two discrete probability distributions $P,Q$ with domain $\mathcal{X}$ is given by 

$$ H^2(P,Q) = \frac{1}{2}\sum_{\mathcal{X}}\left(\sqrt{P(x)} - \sqrt{Q(x)}\right)^2$$.

Here and elsewhere in the proof we will use $\sum_{\mathcal{X}}$ as shorthand for $\sum_{x\in\mathcal{X}}$.

Now we will detail the series of games:

\paragraph{$\mathcal{G}_0$ (Non-private ecosystem).}
This game is the non-private ecosystem, so $\advantage{0}{\adv} = \advantage{\targetingfunction,I,\browsingfunction,\engagementfunction,\attributionfunction,I, n}{\mathcal{A},\mathcal{D}_0,f_s} > \frac{2}{3}$.

Let $k$ be the number of $\texttt{Report}$ messages that $\Adv$ sends to $\idealmetricsfull$. Then, the output of metrics is a series of reports $\hat{r} = \{r_1,...,r_k\}$. 
Let $f_s$ be the processing function that the adversary $\mathcal{A}_1$ is using to distinguish over. \footnote{To provide some practical intuition, one could imagine that $f_s$ is akin to Fisher's exact test being used to perform A/B testing on the reports outputted by metrics (which could be thought of as counting the number of clicks per ad).} Thus, distinguishing on this output would indicate both that:
\begin{enumerate}
    \item There is a difference in the underlying distribution of user features in $\Dzero$ and $\Done$, and
    \item This difference corresponds to the features targeted by the two ads that are being A/B tested.
\end{enumerate}
From the first observation that $\Dzero$ and $\Done$ are different, Fact \ref{fact:npbound} gives us that the Hellinger distance characterizes the sample complexity $SC^{\Dzero,\Done}=\Theta(1/H^2(\Dzero,\Done))$ between two distributions.
From the second observation that the statistical test $f_s$ is measuring some specific property of the distributions (i.e., A/B testing on a feature), we find that the distributions must differ in regards to that particular property.

\paragraph{$\mathcal{G}_1$ (Switch to $\mathcal{A}'=(\mathcal{A}_0,\Adv',\mathcal{A}_1')$):}
In this game, we consider an adversary who receives the \emph{input} of Metrics rather than the output. This is essentially a data processing inequality: anything that is performed by Metrics can instead be performed by the adversary itself if it is helpful in its distinguishing role.

In more detail, consider a modified $\Adv$' that does not interact with $\idealmetricsfull$, and uses the input to the metrics functionality as its output. In particular, the output distribution from $\Adv$' is a function of the total output from  $\idealengagementfull$. Now, we will show why $\mathcal{A}_1$ implies we can construct a $\mathcal{A}_1'$ that can distinguish here. If the output of $\idealmetricsfull$
is distinguishable, then its input must also be distinguishable with advantage at least $\advantage{0}{\adv}$. Note that the input to metrics is the output of engagement, which is a list of browsing histories $\{(\texttt{site},\texttt{ad},\texttt{conversion})\}$ for each user. Formally, we can say that for $n$ users, the output of engagement is $\{\texttt{browsing-history}_1,...,\texttt{browsing-history}_n\}$, where for user $i$ with features $\texttt{features}_i$:
\begin{multline*}
    \texttt{browsing-history}_i \\ = \{(\texttt{site}_{i,1}, \texttt{ad}_{i,1}, \engagementfunction(\texttt{features}_i,\texttt{site}_{i,1}, \texttt{ad}_{i,1})),...\}
\end{multline*}
where
$$
\texttt{ad}_{i,1}=\targetingfunction(\texttt{active-campaigns}, \texttt{features}_i, \texttt{site}_{i,1}).
$$

Additionally, we note that for these to result in a distinguishable set of reports, there must be some pair:
$$(\texttt{audience}_1, \texttt{ad}_1), (\texttt{audience}_2, \texttt{ad}_2) \in \texttt{active-ads}$$ that had distinguishable metrics.
Since metrics simply performs post-processing on the result of engagement, there exists an adversary $\mathcal{A}_1'$ that can distinguish this pair with advantage $\advantage{1}{\adv} \geq \advantage{0}{\adv}$.

\paragraph{$\mathcal{G}_2$ (Add $(f_t',\rho)$):}
Targeting and Engagement utility mean that some information about the difference in the two distributions is still leaked, and we can use distribution testing to distinguish here. So, in this game, we replace $(f_t,I)$ with $(f_t',\rho$), where $f_t'$ is $\alpha_t'$ useful with respect to $close$ and $\rho$. As in the previous game, we focus on the task of $\mathcal{A}^2$ being able to distinguish the combined output of $\idealengagementfull$ with respect to the pair: $$(\texttt{audience}_1, \texttt{ad}_1), (\texttt{audience}_2, \texttt{ad}_2) \in \texttt{active-ads}.$$

In particular, we want to show that by expanding the campaign size, we can construct an adversary that will be able to still distinguish with the same advantage. We leverage the fact that the size of the campaign determines the number of samples pulled from the distribution $\mathcal{D}_b$. Thus, by finding the sample complexity for distinguishing between the output distributions, we can determine the necessary campaign size. We note that the adversary could in theory construct a small campaign or one that is not relevant to the difference between the distributions, but our requirement that the adversary is able to distinguish eliminates the need to consider such scenarios.

We note that advertisers will try to use $\idealtargetingfull$ to deliver an ad dependent on a particular test feature. In this case, the addition of $(f_t',\rho)$ continues to allow targeting based on this feature, however, just less accurately (our analysis also holds for the case where $\alpha_t'=0$). Formally, for the pair ($\texttt{audience}_1, \texttt{ad}_1), (\texttt{audience}_2,$ $ \texttt{ad}_2)$ and a feature vector representing a user $x$, if 
$$
    close(\texttt{audience}_1, x) - close(\texttt{audience}_2, x) = \close_x,
$$
then:
\begin{multline*}
        \Pr\left[\texttt{ad}_1 \leftarrow \targetingfunction(\texttt{active-campaigns}, \rho(x),  \texttt{site})\right]  \\
        - \Pr\left[\texttt{ad}_2 \leftarrow \targetingfunction(\texttt{active-campaigns}, \rho(x), \texttt{site})\right]  \geq \alpha_t' \cdot \close_x,
    \end{multline*}
where $\alpha_t'< \alpha_t$.

We note that the size of the campaign that is distinguishable in the non-private setting is $n$ with success probability $\frac{5}{6}$ (since the advantage is $\frac{2}{3}$). Additionally, the distribution of the output of engagement is the distribution of ad conversions. In particular, we can look at the conversions for ($\texttt{audience}_1, \texttt{ad}_1)$ in comparison to $(\texttt{audience}_2,$ $ \texttt{ad}_2)$. We express this by 
\begin{align*}
    R_b(x)&=\Pr[x~sampled~from~\mathcal{D}_b]\Pr[show~x~\texttt{ad}_1]\Pr[x~clicks~on~\texttt{ad}_1] \\ 
    &= \mathcal{D}_b(x)\cdot(\frac{1+\alpha_t\cdot\close_x}{2})\cdot(\alpha_e\cdot close(\texttt{ad}_1, x)).
\end{align*}  

From Fact \ref{fact:npbound} we know $n=\Theta(1/H^2(R_0,R_1))$ and in particular from Facts \ref{fact:nplower} and \ref{fact:npupper} that 

$$\frac{\textup{ln}(\frac{3}{2})}{4\cdot H^2(R_0,R_1)} < n < \frac{1}{H^2(R_0,R_1)}.$$

For the private case, we have the updated distributions for the output of engagement: 
\begin{align*}
    R_b'(x)&=\Pr[x~sampled~from~\mathcal{D}_b]\Pr[show~x~\texttt{ad}_1]\Pr[x~clicks~on~\texttt{ad}_1] \\ 
    &= \mathcal{D}_b(x)\cdot(\frac{1+\alpha_t'\cdot\close_x}{2})\cdot(\alpha_e\cdot close(\texttt{ad}_1, x)).
\end{align*}  

Therefore, the campaign size needed for the adversary to distinguish would be

$$\frac{\textup{ln}(\frac{3}{2})}{4\cdot H^2(R_0',R_1')} < n_t' < \frac{1}{H^2(R_0',R_1')}.$$

Note that $\Dzero,\Done,R_0,R_1,R_0',R_1'$ all have the same domain $\mathcal{X}$, which is the universe of potential users.

In order to compute the bound on expansion factor $z = n_{t'}/n$ we must take an upper bound for the private case and a lower bound for the non-private case. Let $\gamma = \frac{4}{\textup{ln}(\frac{3}{2})}$, then

\begin{multline*}
    z = \frac{\gamma H^2(R_0,R_1)}{H^2(R_0',R_1')} \\
    = \frac{\gamma\frac{1}{2} \sum_{\mathcal{X}} \left( \sqrt{R_0(x)} - \sqrt{R_1(x)} \right)^2}{\frac{1}{2} \sum_{\mathcal{X}} \left( \sqrt{R_0'(x)} - \sqrt{R_1'(x)} \right)^2} \\
    = \gamma\frac{1}{2} \sum_{\mathcal{X}} \left( \sqrt{\mathcal{D}_0(x) \cdot \left( \frac{1 + \alpha_t \cdot \close_x}{2} \right) \cdot \left( \alpha_e \cdot \text{close}(\texttt{ad}_1, x) \right)} - \right. \\
    \left. \sqrt{\mathcal{D}_1(x) \cdot \left( \frac{1 + \alpha_t \cdot \close_x}{2} \right) \cdot \left( \alpha_e \cdot \text{close}(\texttt{ad}_1, x) \right)} \right)^2 \\
    / \frac{1}{2} \sum_{\mathcal{X}} \left( \sqrt{\mathcal{D}_0(x) \cdot \left( \frac{1 + \alpha_t' \cdot \close_x}{2} \right) \cdot \left( \alpha_e \cdot \text{close}(\texttt{ad}_1, x) \right)} - \right. \\
    \left. \sqrt{\mathcal{D}_1(x) \cdot \left( \frac{1 + \alpha_t' \cdot \close_x}{2} \right) \cdot \left( \alpha_e \cdot \text{close}(\texttt{ad}_1, x) \right)} \right)^2 \\
    = \frac{\gamma \alpha_e \sum_{\mathcal{X}} \text{close}(\texttt{ad}_1, x) (\frac{1 + \alpha_t \cdot \close_x}{2})\left( \sqrt{\mathcal{D}_0(x)} - \sqrt{\mathcal{D}_1(x)} \right)^2}{\alpha_e \sum_{\mathcal{X}} \text{close}(\texttt{ad}_1, x)(\frac{1 + \alpha_t' \cdot \close_x}{2}) \left( \sqrt{\mathcal{D}_0(x)} - \sqrt{\mathcal{D}_1(x)} \right)^2} \\
    = \frac{\gamma\sum_{\mathcal{X}} \text{close}(\texttt{ad}_1, x) (1 + \alpha_t \cdot \close_x)\left( \sqrt{\mathcal{D}_0(x)} - \sqrt{\mathcal{D}_1(x)} \right)^2}{\sum_{\mathcal{X}} \text{close}(\texttt{ad}_1, x)(1 + \alpha_t' \cdot \close_x) \left( \sqrt{\mathcal{D}_0(x)} - \sqrt{\mathcal{D}_1(x)} \right)^2} \\
    = \gamma\frac{A+\alpha_tB}{A+\alpha_t'B} = \gamma\frac{1+\alpha_tK}{1+\alpha_t'K}.
\end{multline*}

Thus, we have $n_{t'} < n \cdot \gamma\frac{1+\alpha_tK}{1+\alpha_t'K} $ where 
$$A = \sum_{\mathcal{X}} \text{close}(\texttt{ad}_1, x) \left( \sqrt{\mathcal{D}_0(x)} - \sqrt{\mathcal{D}_1(x)} \right)^2$$, $$B = \sum_{\mathcal{X}} \text{close}(\texttt{ad}_1, x) \close_x\left( \sqrt{\mathcal{D}_0(x)} - \sqrt{\mathcal{D}_1(x)} \right)^2$$ and $K = \frac{B}{A}$. 

As shown above we computed this ratio by simplifying down the expression $$\frac{\gamma\frac{1}{2} \sum_{\mathcal{X}} \left( \sqrt{R_0(x)} - \sqrt{R_1(x)} \right)^2}{\frac{1}{2} \sum_{\mathcal{X}} \left( \sqrt{R_0'(x)} - \sqrt{R_1'(x)} \right)^2}.$$
We did this by expanding out the terms defining these distributions and identifying that the differences in the numerator and denominator both had like terms that could be factored out. Many of these like terms then canceled out between the numerator and denominator and we were left with a relatively simple expression of the ratio.

By looking at how the expression $\frac{1+\alpha_tK}{1+\alpha_t'K}$ grows with $K$ and noting that $0\leq K$ we can find additional bounds on $n_t'$. In particular, $1\leq \frac{1+\alpha_tK}{1+\alpha_t'K} < \frac{\alpha_t}{\alpha_t'}$ so we have

$$n\cdot\gamma < n_{t'} < n \cdot \gamma\frac{\alpha_t}{\alpha_t'}. $$

The lower bound tells us that when $K$ approaches 0, which indicates that $B<<A$ (there is a very small difference in the closeness of users to the two audiences) then this ratio is dominated by $\gamma$. However, the upper bound tells us that when $A<<B$ (there is a very large difference in the closeness of users to the two audiences) then this ratio is dominated by $\frac{\alpha_t}{\alpha_t}$. Thus, private targeting mechanisms make the most difference when there is a large difference in how close users are to the two potential audiences in question. 

As a result, there exists an adversary can distinguish by increasing the number of samples (campaign size) to $n_{t'}$ to distinguish. They can use their knowledge of $\Dzero, \Done, \alpha_t, \alpha_t', \rho, \alpha_e$ to compute this value. 
Thus $\advantage{2}{\adv} = 2/3 = \advantage{1}{\adv}$.

\paragraph{$\mathcal{G}_3$ (Distinguish on metrics again):}
In this game, we use the fact that metrics doesn't do anything currently, so it does not impact the adversary's advantage. Hence, we will switch back to interacting with $\Adv$, whose output is a function of the output of $\idealmetricsfull$, where we are still using $I$ as our report creation function. Thus, like in the non-private case, reporting is transparently forwarding through the results from engagement. Thus, $\advantage{3}{\adv} = \advantage{2}{\adv}$.

\paragraph{$\mathcal{G}_4$ (Add $f_r^\epsilon$):}
In this final game, we leverage the fact that metrics has utility, and therefore being able to still perform statistical tests in the private setting means the differences in the distributions remain detectable and we can use differentially-private distribution testing techniques to distinguish here. In more detail, we now use the reporting function $\reportcreationfunction^\epsilon$ (with respect to $f_s$) instead of the identity $I$. Because $\reportcreationfunction^\epsilon$ is $\alpha_r$-utility-preserving with respect to $f_s$, this means that the impact of the reporting function on the statistical test cannot be more that $\alpha_r$. As mentioned in \cref{defn:utility:metrics}, $\alpha_r$ can be thought of as a bound on the change in the error rate for $f_s$.

Like in our previous games, our goal here is to show that by amplifying the size of the campaign, we can still allow for the adversary to distinguish with a noticeable advantage.
Crucially, in order to use this to find the necessary sample complexity, we need to show that the two distributions are not the same, as otherwise the result will be undefined.

We know that the underlying $\Dzero$ and $\Done$ distributions are different and that from $\mathcal{G}_2$ that this difference is distinguishable on the output of engagement. Thus, the question is whether the reporting function may completely flatten this difference. However, we know that $f_r^\epsilon$ is bounded in how much it can impact the result of the statistical test $f_s$ that is performed on the data. Since this test is measuring a property that is different between $\Dzero$ and $\Done$ (observation (2) from $\mathcal{G}_0$), that means that $f_r^\epsilon$ cannot completely flatten this difference. Thus, the distributions of interest here: $f_r^\epsilon(R_0')$ and $f_r^\epsilon(R_1')$ must have a non zero difference. (This is a slight abuse of notation but $f_r^\epsilon(R_b')$ is the final distribution from applying the reporting function to the engagement output distribution $R_b'$.) 

We can make use of Fact \ref{fact:private} which gives us that the sample complexity for distingushing between these distributions $f_r^\epsilon(R_0'),f_r^\epsilon(R_1')$  $SC_{\epsilon}^{P,Q}$ is $n_t' \leq SC_{\epsilon}^{R_0',R_1'} \leq \frac{10n_t'}{\epsilon}$ with advantage $2(\frac{4}{5}(\frac{\advantage{3}{\adv}+1}{2})+\frac{1}{10})-1 = \frac{8}{10}\advantage{3}{\adv}$.

Thus, we can apply this and set the amplified campaign size to $n_{r} = \frac{10n_t'}{\epsilon} < \frac{10\frac{4}{\textup{ln}(\frac{3}{2})} n}{\epsilon} \cdot n \cdot \gamma\frac{1+\alpha_tK}{1+\alpha_t'K} $. This value is computable with the information used to construction $R_b'$ as well as the $\epsilon$ parameter used in reporting. With this campaign size we have a distinguishing advantage $\advantage{4}{\adv} = \frac{8}{10}\advantage{3}{\adv}$.

The overall lemma follows by combining the bounds on each pair of adjacent games. This gives us a distinguishing advantage $p'=\advantage{4}{\adv} =\frac{8}{10}\advantage{3}{\adv}=\frac{8}{10}\frac{2}{3}= \frac{8}{15} = p - \frac{2}{15}$.

\section{Proof of \cref{thm:distinguishingimpliesnoprivacy}}

We will now show why any ad ecosystem where an adversary $\mathcal{A}=(\mathcal{A}_0,\Adv,\mathcal{A}_1)$ has advantage $\advantage{\targetingfunction,\rho,\browsingfunction,\engagementfunction,\attributionfunction,\reportcreationfunction, n}{\mathcal{A},\mathcal{D}_0, f_s} > 0$ in distinguishing between two distributions $\mathcal{D}_0$ and $\mathcal{D}_1$, means that this ad ecosystem cannot achieve Dataset Attribute Privacy as defined in \cref{defn:datasetattributeprivacy}. In this case, the audience of users is a set of feature vectors, where each attribute is just a single bit that is 1 if the user has that feature.

Assume for the sake of contradiction that this ad ecosystem has dataset attribute privacy where the private function $g(X_i)$ is a sum of the values of the $i^\text{th}$ feature. Also, let $f_s$ be the summary statistic $F$, which reveals how many users in the dataset had that feature.

From the fact that a distinguisher $\mathcal{A}$ exists, we know that the output of the environment is distinguishable. This directly implies that the output of $\idealmetricsfull$ is distinguishable between the two probability distributions $\mathcal{D}_0$ and $\mathcal{D}_1$. In particular, if $k$ was the total number of reports that $\Adv$ requested from $\idealmetricsfull$, the output it distinguishes on is a list of reports $\hat{r}=\{r_1,...,r_k\}$. Let $f_s$ be the processing function such that the output of the adversary is the output of $f_s(\hat{r})$. Since, $\hat{r}=\{r_1,...,r_k\}$ is distinguishble, there must be some $r\in \hat{r}$ that is distinguishable.

By \cref{defn:datasetattributeprivacy} if attribute privacy were achieved for all parameters governing the distribution of users in $\idealsocietyfull$ then the output of $f_s(\hat{r})$ should be independent of which distribution $\idealsocietyfull$ sampled users from ($\Dzero$ vs $\Done$). However, $\mathcal{A}$ is able to distinguish on this output, which implies that there must be some report where the result is dependent on choice of distribution. Therefore, there must be some attribute for which attribute privacy is not preserved. 

This is a contradiction to our claim that we had attribute privacy, and thus we conclude that the existence of an adversary that can distinguish between the underlying distributions in an ad ecosystem implies no attribute privacy.

\section{Proof of \cref{thm:utilityimpliesleakage}}

\label{app:theorem-proof}

Our proof of \cref{thm:utilityimpliesleakage} follows from \cref{thm:distinguishing} and \cref{thm:distinguishingimpliesnoprivacy} in a straightforward manner.
Suppose we had an ads ecosystem composed of instantiations of $\idealtargetingfull,$ $\idealengagementfull,$ and $\idealmetricsfull$ that are \emph{useful} (as defined by \cref{defn:utility:targeting,defn:utility:engagement,defn:utility:attribution,defn:utility:metrics}).

By \cref{thm:distinguishing} we see that for such an ad ecosystem, we will have an adversary $\mathcal{A}$ that can distinguish between $\mathcal{D}_0$ and $\mathcal{D}_1$ with probability $\frac{8}{15}$. By, \cref{thm:distinguishingimpliesnoprivacy} we see that if such a distinguishing adversary exists, then this ad ecosystem will not satisfy attribute privacy for some attribute of that campaign's audience.

Thus, we conclude that any useful ads ecosystem
for a given ad campaign will not satisfy attribute privacy for some attribute of that campaign's audience.

\section{Extended Empirical Evaluation}
\label{app:empirical-plots}

\begin{figure}[t]
\includegraphics[width=0.45\textwidth]{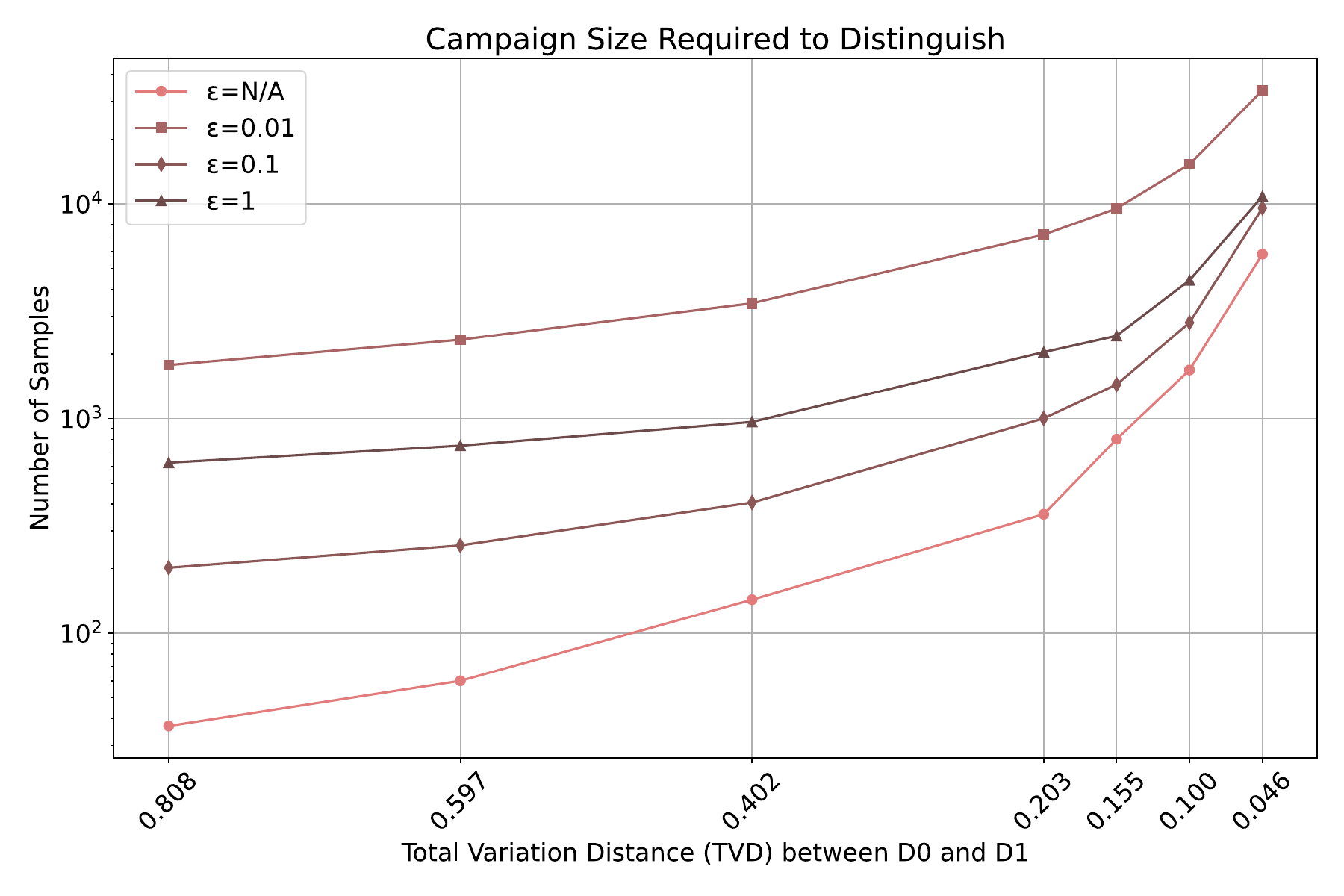}
\caption{Impact of $\epsilon$ on sample complexity.}\label{fig:empirical-distinguishing-epsilon}
\end{figure}

As shown in \cref{fig:empirical-distinguishing-epsilon}, lower $\epsilon$ values increase the amount on noise added to the differentially-private metrics which reduces the confidence of the distinguisher at low sample sizes. As the number of samples increases, the signal to noise ratio improves and the sample complexity of the differentially-private lines approaches the baseline.

\begin{figure}[t]
\includegraphics[width=0.45\textwidth]{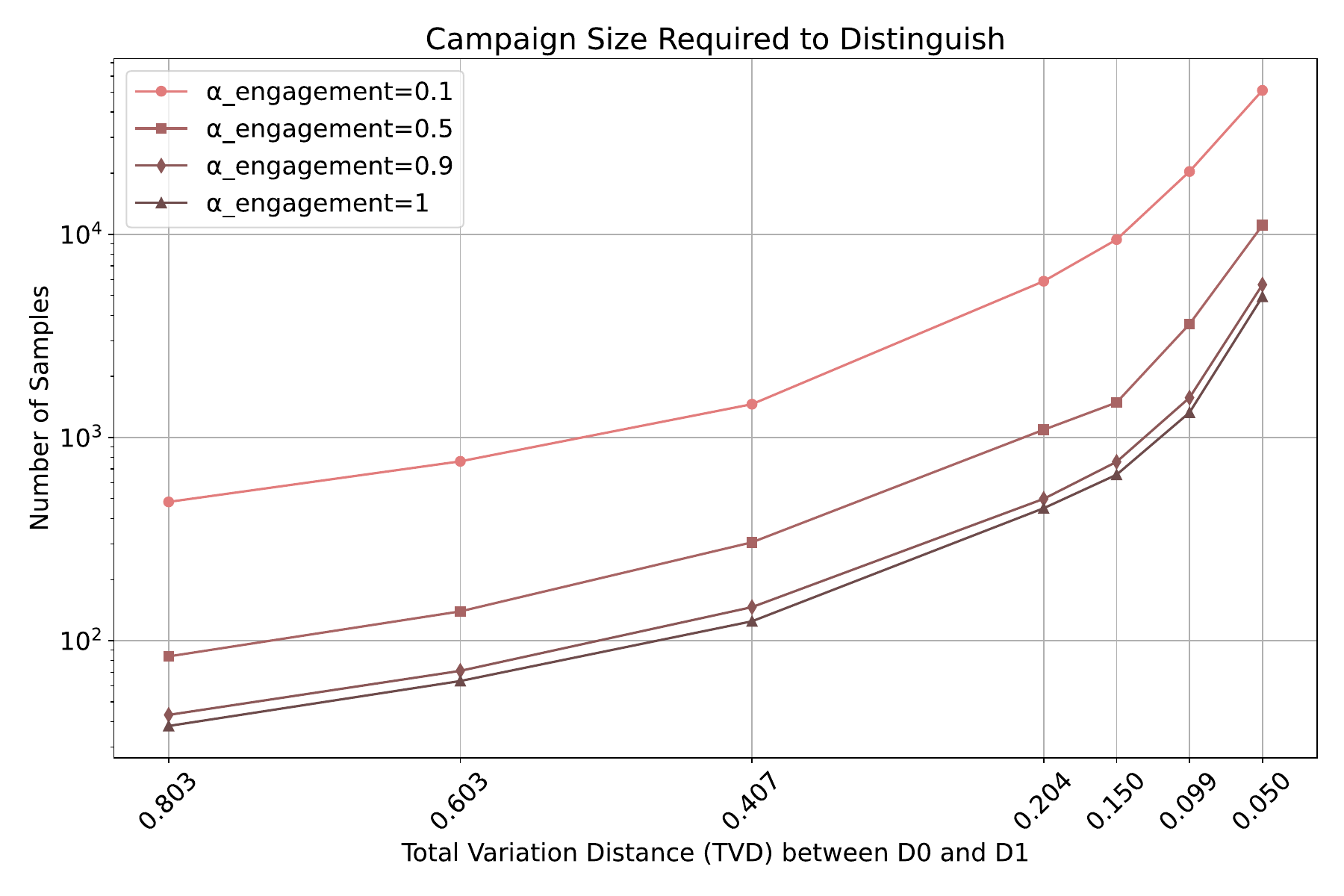}
\caption{Impact of $\alpha$-engagement on sample complexity.}\label{fig:empirical-distinguishing-engagement}
\end{figure}

$\alpha$-engagement impacts the overall click through rate so reducing this value reduces the overall click probability as shown in \cref{fig:empirical-distinguishing-engagement}. Consequently, this reduces the absolute difference in click probability between $\mathcal{D}_0$ and $\mathcal{D}_1$ making distinguishing more challenging. Generally, ad campaigns have very low click through rates (less than a percent) so low $\alpha$-engagement values are the most realistic.

\begin{figure}[t]
\includegraphics[width=0.45\textwidth]{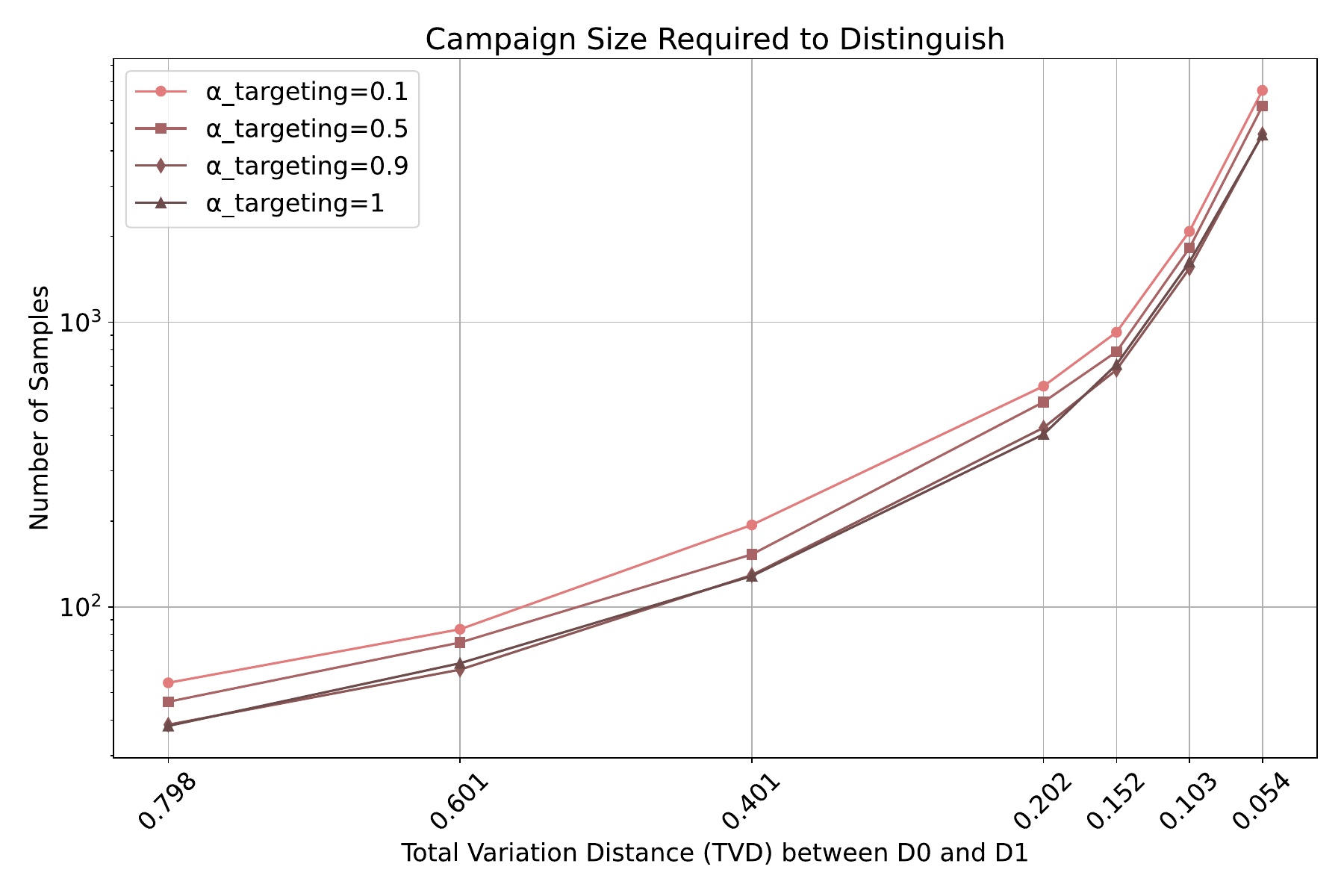}
\caption{Impact of $\alpha$-targeting on sample complexity.}\label{fig:empirical-distinguishing-targeting}
\end{figure}

\cref{fig:empirical-distinguishing-targeting} depicts how $\alpha$-targeting has a less strong impact on overall sample complexity in distinguishing than either $\alpha$-engagement or the $\epsilon$ value for differentially-private metrics.
This is largely due to the way we use it as a modifier for $close()$ in the targeting utility definition---most users are not a perfect match for the ad feature vector and may be relatively close to both ads causing a small difference in probability of preferring one over the other even when $\alpha$-targeting is 1. Alternative designs for targeting utility could have $\alpha$-targeting instead indicate how likely targeting is to prefer the closer ad, regardless of \textit{how} close it was and in this case $\alpha$-targeting would have a stronger impact. 

Additionally, we might expect the audience distribution and definition of $close()$ to affect how impactful $\alpha$-targeting is. E.g., if the audience is well-specified and the users tend to share similar features aside from the test bit, then $close()$ might decide to weight that bit more strongly than the others when deciding which ad is more relevant.

\section{Background on Pufferfish Privacy}
\label{app:pufferfish}

The Pufferfish privacy framework~\cite{kifer2014pufferfish} is intended to capture scenarios where privacy is desired for sensitive data, and, moreover, the sensitive data might be correlated with some of the other features in the dataset. Whereas differential privacy treats all data as sensitive and therefore requires hiding all correlations, Pufferfish privacy is more flexible and allows for revealing some features but only up to the point that sensitive data remains hidden. Consider for example genetic traits or the transmission of disease: information about the hair color of a person’s family members can allow for inferences of their own hair color and knowing that many people in a person’s community have the flu is revealing of that person’s health data. Resolving this under standard DP typically involves considering these correlated groups of people as a single entry. While this is effective, it introduces unmanageable levels of noise as the group size grows.

The Pufferfish framework allows for differentially-private statistics on correlated data without assuming that all entries are fully correlated. This allows it to achieve protection for correlated data at a more manageable level of noise. The framework consists of three parts:

\begin{enumerate}
\item A set of secrets $\mathcal{S}$: this is the information that should \textbf{not} be revealed (or inferrable) by any output statistics. E.g., from the examples above, the secrets would be the flu status of any individual or their hair color.
\item Pairs of potential secrets $\mathcal{Q} = (\mathcal{S} \times \mathcal{S})$: these are the values of secrets that should be indistinguishable given the output. For example: perhaps it should not be possible to distinguish between ``Alice has the flu’’ and ``Alice is healthy,’’ or to distinguish between any combination of possible hair colors for Alice
(e.g., brown vs.\ green hair, brown vs.\ blonde hair, etc).
This list should be viewed as a denylist: any two events X and Y that do \emph{not} form a pair (X, Y) in the list are (implicitly) allowed to be distinguishable by the adversary.
\item A set of distributions $\Theta$ that could plausibly generate the dataset: $\Theta$ defines the correlations between the individual datapoints in the dataset. E.g., different $\Theta$ could specify varying levels of contagiousness for the flu or different probabilities that an individual has a certain hair color given the hair colors of their family members.
\end{enumerate}

\begin{definition}[Definition 2.1 from \citet{song2017pufferfish}]
A privacy mechanism $M$ is said to be $\epsilon$-Pufferfish private in a framework $(\mathcal{S}, \mathcal{Q}, \Theta)$ if for all $\theta \in \Theta$ with $X$ drawn from distribution $\theta$, for all secret pairs $(s_i, s_j) \in \mathcal{Q}$, and for all $w \in \text{Range}(M)$, we have
\begin{equation} \label{eqn:defpf}
e^{-\epsilon} \leq \frac{P_{M,\theta}(M(X)=w|s_i, \theta)}{P_{M,\theta}(M(X)=w|s_j, \theta)} \leq e^{\epsilon} 
\end{equation}
when $s_i$ and $s_j$ are such that $P(s_i|\theta) \neq 0$, $P(s_j | \theta) \neq 0.$
\end{definition}

Attribute privacy uses the Pufferfish framework to focus on privatizing the distribution of sensitive attributes within a dataset. There, the secrets $\mathcal{S}$ are the output of some function $g()$ over that sensitive attribute. Like most parameters in Pufferfish privacy, exactly what $g()$ is will be situational and differ between use cases. Here, we care about the fraction of the audience who possesses the sensitive attribute so $g()$ computes this value.

\end{document}